%% file: B02JpsiKK.tex
\documentclass[12pt,a4paper]{article}

\usepackage{lineno}  
\usepackage{graphicx}  
\usepackage{xspace} 
\usepackage{color}
\usepackage{colortbl}
\usepackage{amsmath} 
\usepackage{ifthen} 
\usepackage{subfigure}

\newboolean{pdflatex}
\setboolean{pdflatex}{true} 
\usepackage{rotating} 
\newboolean{articletitles}
\setboolean{articletitles}{true} 

\newboolean{uprightparticles}
\setboolean{uprightparticles}{false} 

\usepackage{amssymb}
\usepackage{amsfonts}
\usepackage{upgreek} 

\usepackage{hyperref}    
\usepackage[all]{hypcap} 

\input{lhcb} 

\usepackage{mciteplus}
\usepackage{bm}
\usepackage{afterpage}
\usepackage{epsfig,accents}
\begin{document}




\begin{titlepage}

\belowpdfbookmark{Title page}{title}

\pagenumbering{roman}
\vspace*{-1.5cm}
\centerline{\large EUROPEAN ORGANIZATION FOR NUCLEAR RESEARCH (CERN)}
\vspace*{1.5cm}
\hspace*{-5mm}\begin{tabular*}{16cm}{lc@{\extracolsep{\fill}}r}
\vspace*{-12mm}\mbox{\!\!\!\includegraphics[width=.12\textwidth]{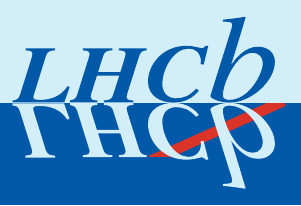}}& & \\
 & & CERN-PH-EP-2013-159\\
 & & LHCb-PAPER-2013-045\\  
 & & \today \\ 
 & & \\
 & & \\
\end{tabular*}

\vspace*{2.0cm}

{\bf\boldmath\Large
\begin{center}
First observation of $\Bdb\to \jpsi \Kp\Km$ and search for $\Bdb\to\jpsi\phi$ decays \\
\end{center}
}

\vspace*{1.0cm}
\begin{center}
\normalsize {
The LHCb collaboration\footnote{Authors are listed on the following pages.}
}
\end{center}

\begin{abstract}
  \noindent
The first observation of the $\Bdb \to \jpsi \KpKm$ decay is presented with a data sample corresponding to an integrated luminosity of 1.0~$\rm fb^{-1}$ of $pp$ collisions at a center-of-mass energy of 7~\tev collected with the \lhcb detector. The  branching fraction is measured to be
$\mathcal{B}(\Bdb \to \jpsi \Kp\Km) = (2.53\pm 0.31 \pm 0.19)\times 10^{-6}$,
where the first uncertainty  is statistical and the second is systematic.  An amplitude analysis of the final state  in the  $\Bdb \to \jpsi \KpKm$ decay is performed to separate  resonant  and  nonresonant contributions in the $\KpKm$ spectrum. Evidence of the $a_0(980)$ resonance is reported with statistical significance of  3.9 standard deviations. The corresponding  product branching fraction is measured to be $\mathcal{B}(\Bdb \to \jpsi a_0(980),~a_0(980)\to\KpKm)=(4.70\pm3.31\pm0.72)\times10^{-7}$, yielding an upper limit of $\mathcal{B}(\Bdb \to \jpsi a_0(980),~a_0(980)\to\KpKm)<9.0\times 10^{-7}$  at  90\% confidence level. No evidence of the resonant decay $\Bdb\to\jpsi\phi$ is found, and  an upper limit on its  branching fraction is set to be $\mathcal{B}(\Bdb \to \jpsi \phi)< 1.9\times 10^{-7}$
at 90\% confidence level.  

\end{abstract}

\vspace{\fill}
\vspace*{2.0cm}

\begin{center}
  Submitted to Phys. Rev. D
\end{center}


{\footnotesize
    \centerline{\copyright~CERN on behalf of the \lhcb collaboration, license \href{http://creativecommons.org/licenses/by/3.0/}{CC-BY-3.0}.}}
\vspace*{2mm}

\clearpage
\newpage
\input{lhcb_author_list}
\end{titlepage}

\renewcommand{\thefootnote}{\arabic{footnote}}
\setcounter{footnote}{0}

\pagestyle{empty}  

\pagestyle{plain} 
\setcounter{page}{1}
\pagenumbering{arabic}



\setcounter{page}{1}
\mbox{~}




\pagestyle{plain} 
\setcounter{page}{1}
\pagenumbering{arabic}


%
\section{Introduction}
\label{sec:Introduction}
\begin{figure}[h]
\centering
\includegraphics[width=0.48\textwidth]{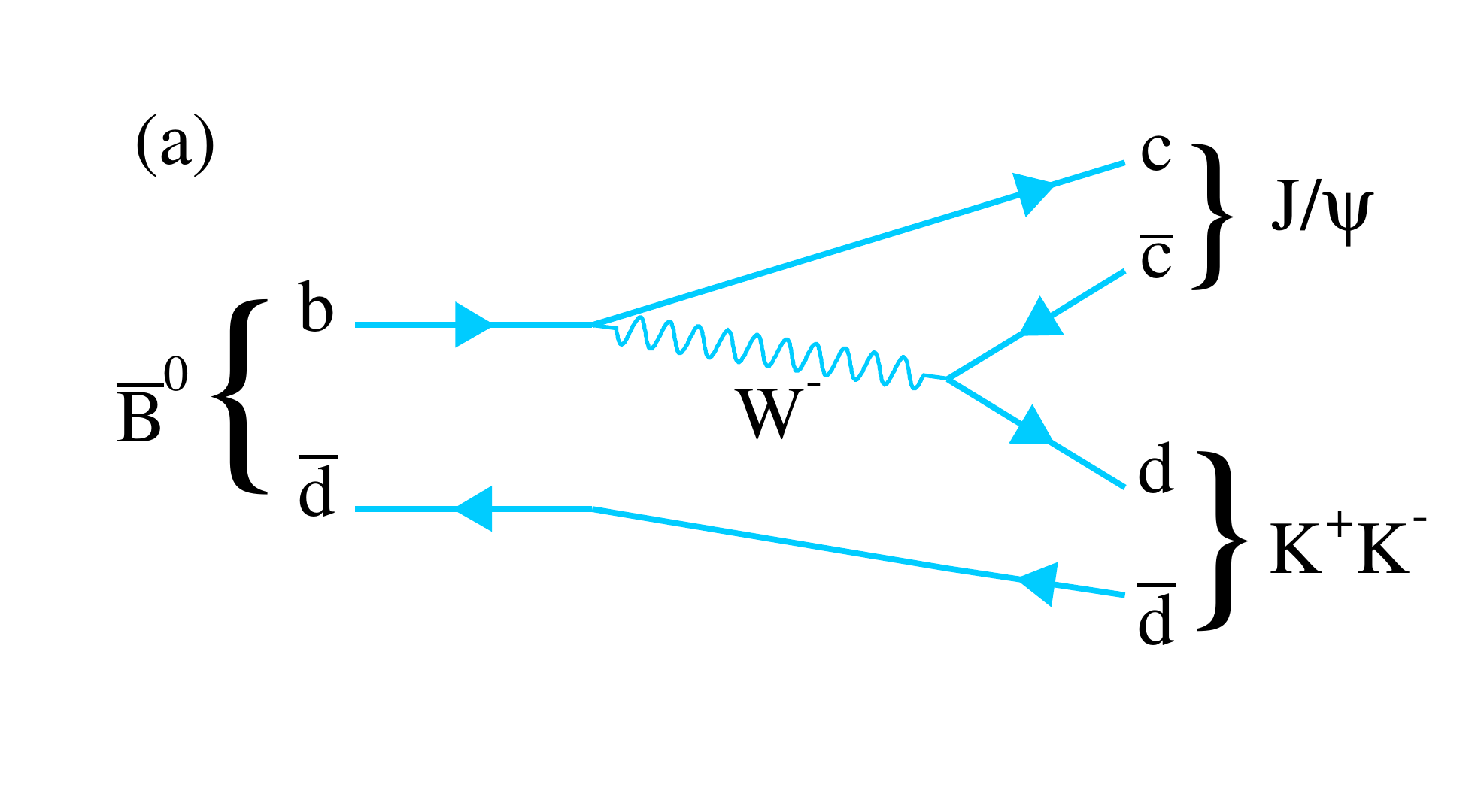}%
\includegraphics[width=0.48\textwidth]{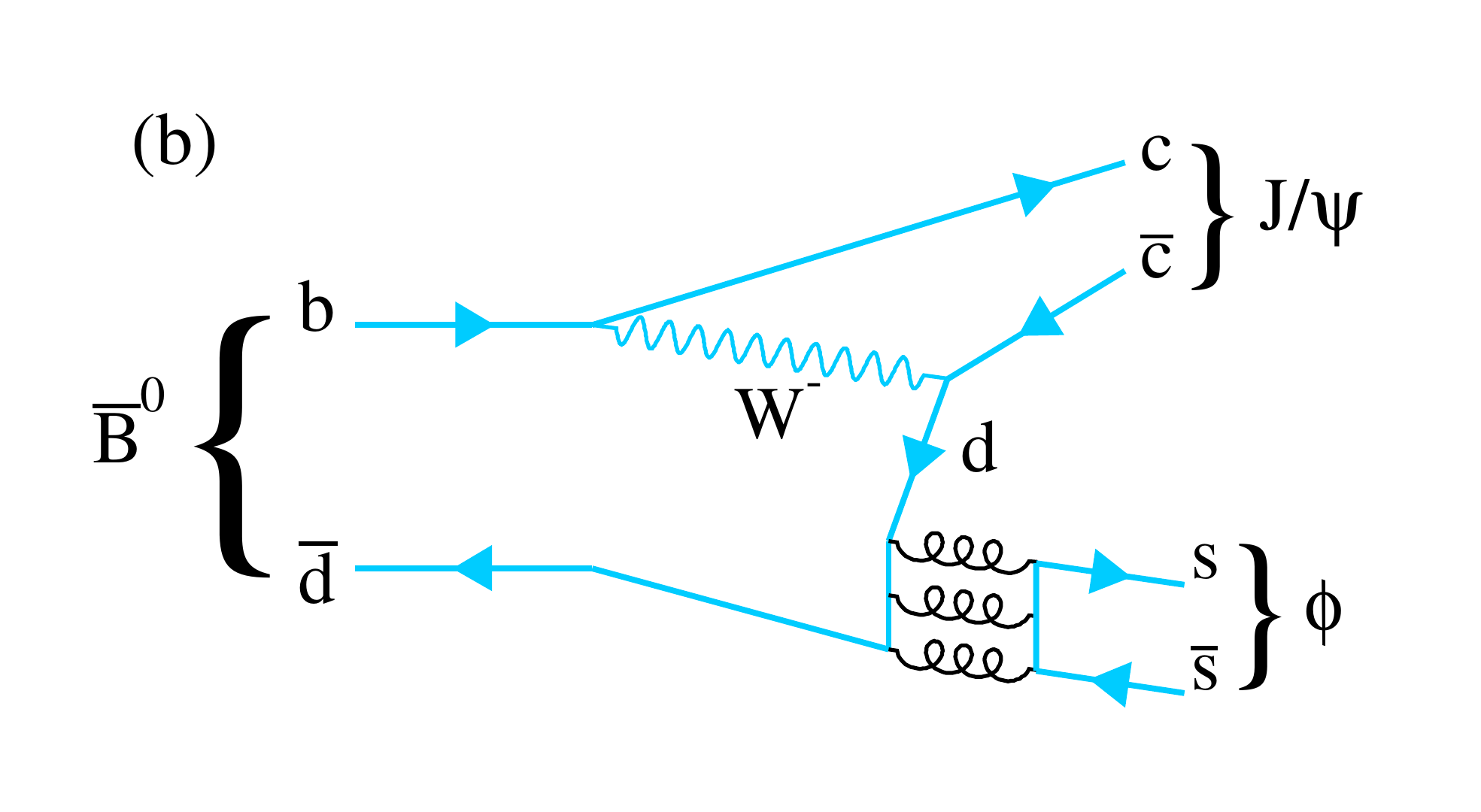}
\vskip -0.5cm
\caption{\small Feynman diagrams for (a) $\Bdb \to \jpsi \Kp \Km$, and (b)  $\Bdb\to \jpsi \phi$.}
\label{fig:feyn1}
\end{figure}

The decays of   neutral $B$ mesons  to a charmonium state and a $ h^+h^-$ pair, where $h$ is either a pion or kaon, play an important role in the study of \CP violation and mixing.\footnote{Charge-conjugate modes are implicitly included throughout the paper.}   
In order to fully exploit these decays for measurements of \CP violation, a better understanding of their final state composition is necessary. Amplitude studies have recently been reported by \lhcb for the  decays $\Bsb \to \jpsi \pip\pim$~\cite{:2012cy}, $\Bsb \to \jpsi \KpKm$~\cite{Aaij:2013orb} and $\Bdb \to \jpsi \pip\pim$~\cite{Aaij:2013zpt}.   Here we perform a similar analysis for $\Bdb \to \jpsi\KpKm$ decays, which are expected to proceed primarily through the Cabibbo-suppressed $b\to \ccbar d$ transition. The  Feynman diagram for the process is shown in Fig.~\ref{fig:feyn1}(a). 
However, the mechanism through which the $\ddbar$ component evolves into a $\KpKm$ pair is not precisely identified.
One possibility is to form a meson resonance that has a $\ddbar$ component in its wave function, but can also decay into $K^+K^-$, another is to excite an $\ssbar$ pair from the vacuum and then have the \ssbar\ddbar system form a $K^+K^-$ pair via rescattering. 
The formation of a $\phi$ meson can occur in this decay either via $\omega-\phi$ mixing which requires  a small $\ddbar$ component in its wave function; or  via  a strong coupling such as shown in Fig.~\ref{fig:feyn1}(b), which illustrates tri-gluon exchange.
Gronau and Rosner predicted  that the dominant contribution is  via $\omega-\phi$ mixing  at the order of $10^{-7}$~\cite{Gronau:2008kk}.

In this paper, we report on a measurement of the branching fraction of the decay $\Bdb\to \jpsi K^+K^-$. A modified Dalitz plot analysis of the final state is performed to study the resonant and nonresonant structures in the $\KpKm$ mass spectrum using the $\jpsi\Kp$ and $\KpKm$ mass spectra and decay angular distributions. This differs from a classical Dalitz plot analysis~\cite{Dalitz} because the $\jpsi$ meson has spin one, so its three helicity amplitudes must be considered. In addition, a search for the decay  $\Bdb\to \jpsi \phi$ is performed.  
\section{Data sample and detector}
The data sample consists of  $1.0~\rm fb^{-1}$ of integrated luminosity collected with the \lhcb detector~\cite{Alves:2008zz} using $pp$ collisions at a center-of-mass energy of 7 TeV.
The \lhcb detector is a single-arm forward
spectrometer covering the \mbox{pseudorapidity} range $2<\eta <5$, designed
for the study of particles containing \bquark or \cquark quarks. The detector includes a high precision tracking system consisting of a
silicon-strip vertex detector surrounding the $pp$ interaction region,
a large-area silicon-strip detector located upstream of a dipole
magnet with a bending power of about $4{\rm\,Tm}$, and three stations
of silicon-strip detectors and straw drift tubes placed
downstream. The combined tracking system has momentum\footnote{We work in units where $c$=1.} resolution
$\Delta p/p$ that varies from 0.4\% at 5\gev to 0.6\% at 100\gev. The impact parameter (IP) is defined as the minimum distance of approach of the track with respect to the primary vertex. For tracks with large transverse momentum, \pt, with respect to the proton beam direction, the IP   resolution is approximately 20\mum. Charged hadrons are identified using two ring-imaging Cherenkov detectors (RICH)~\cite{2013EPJC...73.2431A}. Photon, electron and hadron candidates are identified by a calorimeter system consisting of scintillating-pad and pre-shower detectors, an electromagnetic calorimeter and a hadronic calorimeter. Muons are identified by a system composed of alternating layers of iron and multiwire proportional chambers~\cite{2013JInst...8P2022A}. The trigger consists of a hardware stage, based
on information from the calorimeter and muon systems, followed by a
software stage which applies a full event reconstruction~\cite{Aaij:2012me}.

In the simulation, $pp$ collisions are generated using
\pythia~6.4~\cite{Sjostrand:2006za} with a specific \lhcb
configuration~\cite{LHCb-PROC-2010-056}.  Decays of hadrons
are described by \evtgen~\cite{Lange:2001uf} in which final state
radiation is generated using \photos~\cite{Golonka:2005pn}. The
interaction of the generated particles with the detector and its
response are implemented using the \geant
toolkit~\cite{Allison:2006ve, *Agostinelli:2002hh} as described in
Ref.~\cite{LHCb-PROC-2011-006}.

\section{Event selection}
\label{sec:event-selection}
The reconstruction of $\Bdb \to\jpsi \KpKm$ candidates proceeds by finding $\jpsi\to \mu^+\mu^-$ candidates and combining them with a pair of oppositely charged kaons. 
Good quality of the reconstructed tracks is ensured by requiring the $\chi^2/\rm ndf$ of the track fit to be less than 4, where ndf is the number of degrees of freedom of the fit. To form  a \mbox{$\jpsi\to\mu^{+}\mu^{-}$} candidate, particles identified as muons of opposite charge are required to  have \pt greater than 500~\mev each, and form a vertex with fit $\chi^2$ less than 16. Only candidates with a dimuon invariant mass between $-48$~\mev and $+43$~\mev relative to the observed \jpsi peak are selected, where the r.m.s. resolution is 13.4 MeV.  The requirement is asymmetric due to  final state electromagnetic radiation. The $\mu^+\mu^-$ combinations are then constrained to the \jpsi mass~\cite{PDG} for subsequent use in event reconstruction. 

Each kaon candidate is  required to have \pt greater than 250~\mev and  $\chi^2_{\rm IP}>9$, where the  $\chi^2_{\rm IP}$ is computed as the difference between the $\chi^2$ of the  primary vertex  reconstructed with and without the considered track. In addition, the scalar sum of their transverse momenta, $\pt(\Kp)+\pt(\Km)$, must be greater than 900~\mev. The $\KpKm$ candidates are required to form a  vertex with a $\chi^2$ less than 10 for one degree of freedom. We identify the hadron species of each track  from the difference DLL($h_1-h_2$)  between  logarithms of the likelihoods associated with the two hypotheses $h_1$ and $h_2$, as provided by the RICH detector. Two criteria are used, with the ``loose'' criterion  corresponding to DLL$(K-\pi)>0$, while ``tight'' criterion requires  DLL$(K-\pi)>10$ and DLL$(K-p)>-3$. Unless stated otherwise, we use the tight criterion for the kaon selection.

The \Bdb candidate should have  vertex fit $\chi^2$ less than 50 for five degrees of freedom and a  $\chi^2_{\rm IP}$ with respect to the  primary vertex  less than 25. When more than one primary vertex is reconstructed, the one
that gives the minimum $\chi^2_{\rm IP}$  is chosen. The \Bdb candidate must have a flight distance  of more than 1.5 mm from the associated primary vertex. In addition, the angle between the combined momentum vector of the decay products and the vector formed from the position of the primary vertex to the decay vertex (pointing angle) is required to be smaller than $2.56^{\circ}$. 
\begin{figure}[t]
\centering
\includegraphics[width=0.78\textwidth,height=0.52\textwidth]{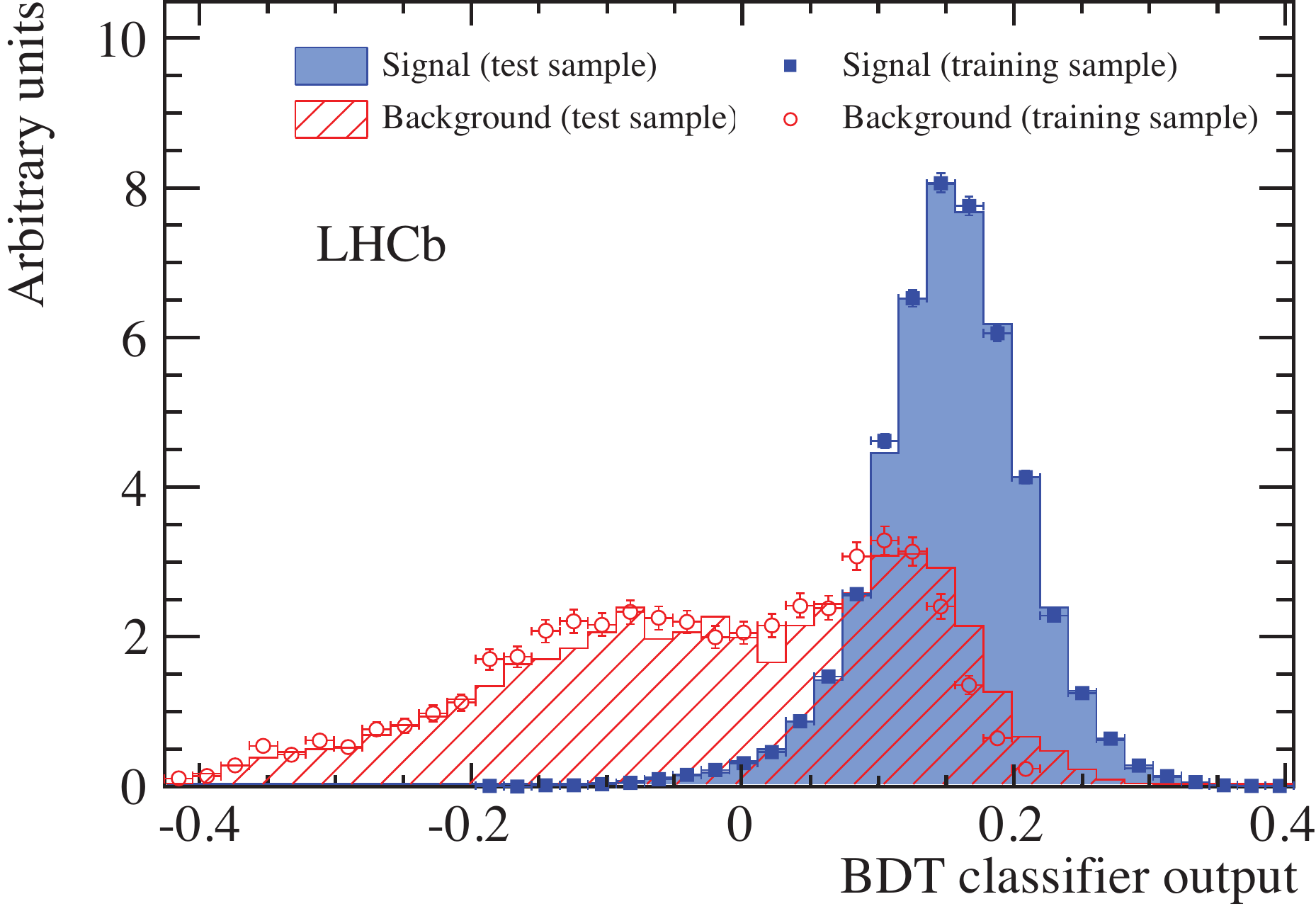}
\vskip -3mm
\caption{\small Distribution of the BDT classifier for both training and test samples of $\jpsi\KpKm$  signal and background events. The signal samples are from simulation and the background samples are from data. The small difference between the background training and test samples is due to the fact that the
sidebands used in the two cases are not identical.} 
\label{fig:BDT_response}
\end{figure}

Events satisfying the above criteria are  further filtered using a multivariate classifier based on a Boosted Decision Tree (BDT) technique~\cite{Breiman}. The BDT uses six variables that are chosen  to provide  separation between signal and background. The BDT variables are
the minimum DLL($\mu-\pi$) of the $\mu^+$ and $\mu^-$, the minimum $\pt$ of the $K^+$ and $K^-$,
the minimum of the  $\chi^2_{\rm IP}$ of the $K^+$ and $K^-$,
the $\Bdb$ vertex $\chi^2$,
the $\Bdb$ pointing angle, and the $\Bdb$ flight distance. The BDT is trained on a simulated sample of $\Bdb\to\jpsi\KpKm$ signal events and a background data sample from the sideband $5180<m(\jpsi\KpKm)<5230$~\mev of the \Bdb signal peak. The BDT is then tested on independent  samples.  The  distributions of the output of the BDT classifier for signal and background are shown in Fig.~\ref{fig:BDT_response}.  
The final selection is optimized  by maximizing $N_S/\sqrt{(N_S+N_B)}$, where  the  expected signal yield $N_S$  and  the expected  background yield $N_B$  are estimated  from the yields before applying the BDT, multiplied by  the efficiencies  associated to various  values of the BDT selection as determined in  test samples. The optimal selection is found to be  $\rm BDT >0.1$, which has an $86\%$ signal efficiency and a  $72\%$  background rejection rate.
\begin{figure}[t]
\centering
\includegraphics[width=0.78\textwidth]{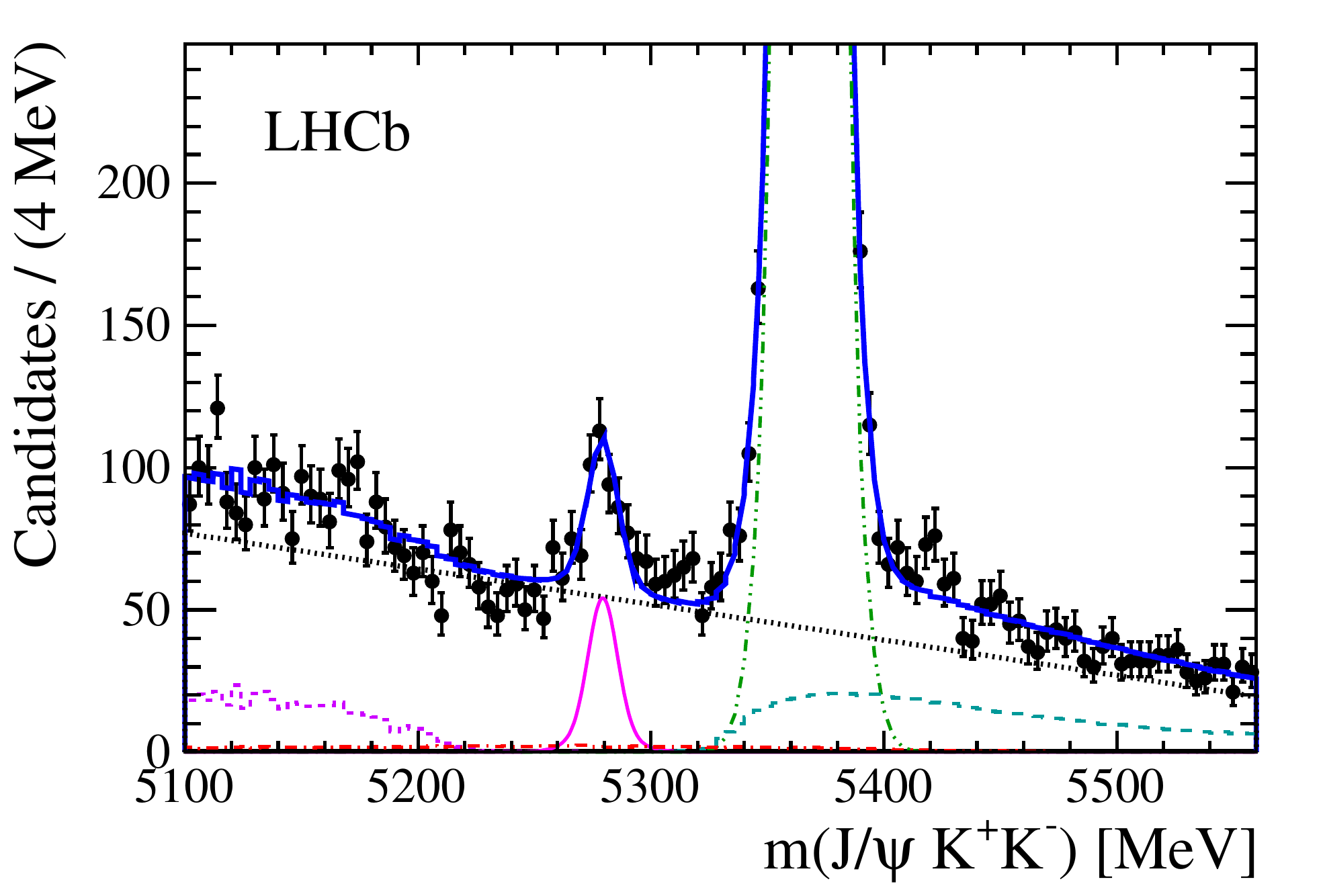}
\vskip -3mm
\caption{\small Invariant mass of $\jpsi \KpKm$ combinations. The  data are fitted with a sum of two Gaussian functions for each signal peak and several background components. The (magenta) solid double-Gaussian function centered at 5280 MeV is the $\Bdb$ signal, the (black) dotted curve shows the combinatorial background, the (green) dashed-dot-dot curve shows the contribution of $\Bsb\to\jpsi \KpKm$ decays, the (violet) dashed shape is the $\Bsb \to \jpsi \Kp\pi^0\Km$ background,  $\Lb\to \jpsi p\Km$ and $\Bdb\to \jpsi\Km\pip$ reflections are shown by (red) dot-dashed and (cyan) long dashed shapes, respectively and the (blue) solid curve is the total. }
\label{fig:B2JpsiKK}
\end{figure}

The invariant mass distribution of the selected $\jpsi \Kp\Km$ combinations is shown in Fig.~\ref{fig:B2JpsiKK}. Signal peaks are observed at both the $\Bsb$ and $\Bdb$ masses  overlapping a smooth   background.  We model the $\Bsb\to \jpsi\Kp\Km$ signal by a sum of two Gaussian functions with common mean; the mass resolution is found to be 6.2~MeV. The shape of the $\Bdb\to \jpsi\Kp\Km$ signal component is constrained to be the same as that of the $\Bsb$ signal. The background components include the combinatorial background, a contribution from the $\Bsb \to \jpsi \Kp \pi^0 \Km$ decay, and reflections from $\Lb \to \jpsi p\Km$ and $\Bdb \to \jpsi \Km\pip$ decays, where a proton in the former and a pion in the latter are misidentified as a kaon. The combinatorial background is described by a linear function. The shape of the $\Bsb \to \jpsi \Kp \pi^0 \Km$ background is taken from simulation, generated uniformly in phase space, with its yield allowed to vary.   The reflection shapes are also taken from simulations, while the yields are Gaussian constrained  in the global fit to the expected values estimated by measuring the number of $\Lb$ and $\Bdb$ candidates in the control region $25-300$ MeV above the $\Bsb$ mass peak. The shape of the $\Lb\to\jpsi p\Km$ reflection is determined from the  simulation  weighted according to the $m(p\Km)$ distribution obtained in Ref.~\cite{Aaij:2013oha}, while the simulations of $\Bdb\to\jpsi\Kstarzb(892)$ and $\Bdb\to\jpsi \Kstarb_2(1430)$ decays are used to study the shape of the $\Bdb\to\jpsi\Km\pi^+$ reflection. From the fit, we extract $228 \pm 27$ \Bdb signal candidates together with $545\pm 14$ combinatorial background   and $20\pm 4$ $\Lb\to\jpsi p\Km$ reflection candidates within $\pm 20$ MeV of the $\Bdb$ mass peak.

\begin{figure}[t]
\centering
\includegraphics[width=0.78\textwidth]{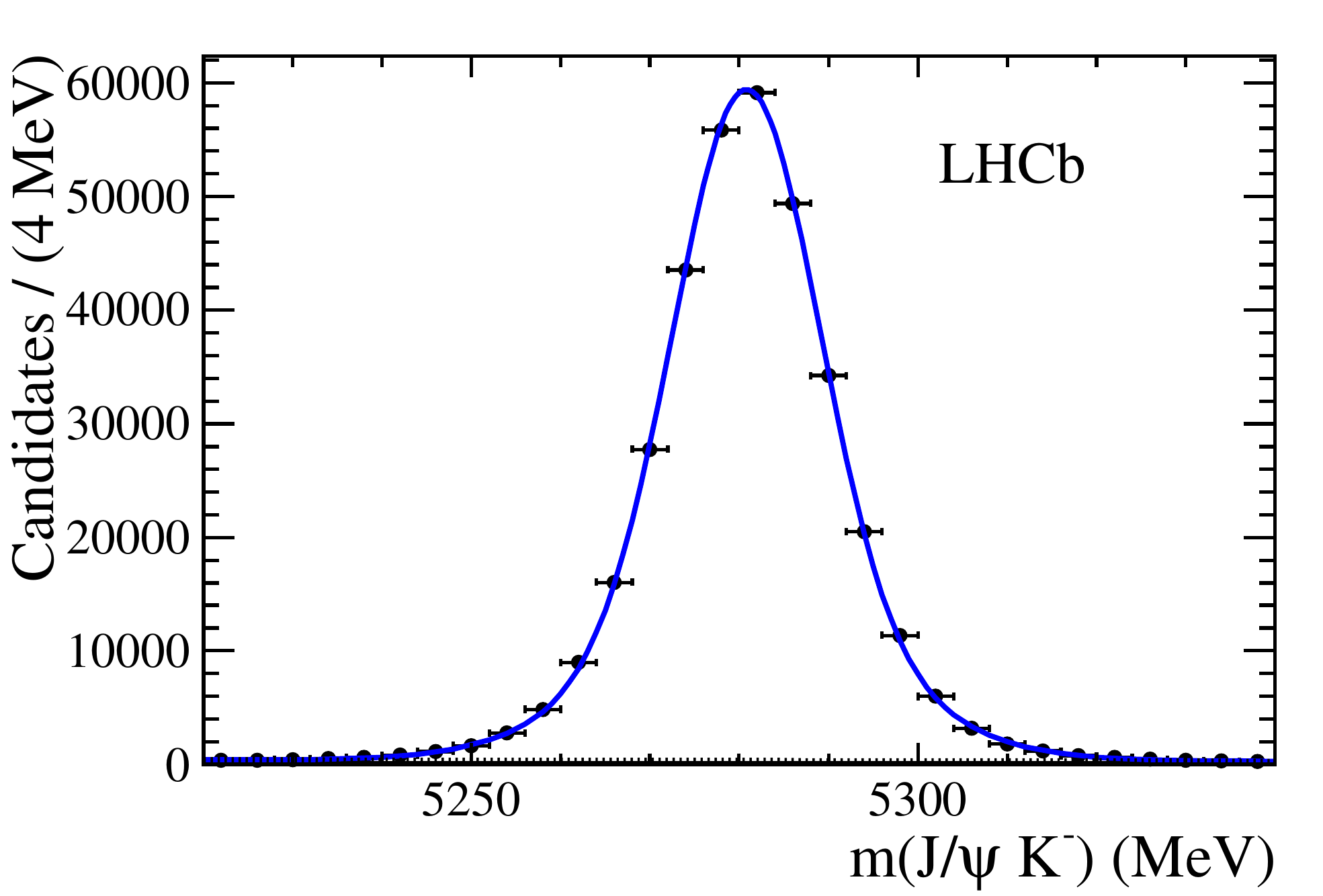}
\vskip -3mm
\caption{\small Fit to the invariant mass spectrum of $\jpsi \Km$ combinations. The (blue) solid  curve is the total and the (black) dotted line shows the combinatorial background.}
\label{fig:Bu2JpsiK}
\end{figure}

We use the decay $\Bm\to \jpsi \Km$ as the normalization channel for branching fraction determinations. The selection criteria are similar to those used for the $\jpsi\KpKm$ final state, except for particle identification requirements since here the loose kaon identification criterion is used. Similar variables are used for the BDT, except that the variables describing the combination of $\Kp$ and $\Km$ in the $\jpsi \KpKm$ final state are replaced by the ones that describe the $\Km$ meson. The BDT training uses $\Bm\to\jpsi\Km$ simulated events as signal and data in the sideband region $5400 <m(\jpsi \Km) < 5450$~\mev as background.   The resulting invariant mass distribution of the $\jpsi \Km$ candidates satisfying BDT classifier output greater than $0.1$ is shown in Fig.~\ref{fig:Bu2JpsiK}. The signal is fit with a sum of two Gaussian functions with common mean and the combinatorial background is fit with a linear function. There are  $322\,696 \pm 596$ signal and $3484 \pm 88$ background candidates  within $\pm20$ MeV of the $\Bm$ peak.

\section{Analysis formalism}
The decay  $\Bdb\to \jpsi K^+K^-$  followed by $\jpsi\to\mu^+\mu^-$ can be described by four variables. These are taken to be the invariant mass squared of $\jpsi K^+$, $s_{12}\equiv m^2(\jpsi K^+)$, the invariant mass squared of $K^+K^-$, $s_{23}\equiv m^2(K^+K^-)$, the $\jpsi$ helicity angle, $\theta_{\jpsi}$, which is the angle of the $\mu^+$ in the $\jpsi$ rest frame with respect to the $\jpsi$ direction in the   $\Bdb$ rest frame, and $\chi$, the angle between the $\jpsi$ and $K^+K^-$ decay planes in the $\Bdb$ rest frame.  Our approach is similar to that used  in the LHCb analyses of   $\Bsb \to \jpsi \pip\pim$~\cite{:2012cy}, $\Bsb \to \jpsi \KpKm$~\cite{Aaij:2013orb} and $\Bdb \to \jpsi \pip\pim$~\cite{Aaij:2013zpt}, where a modified  Dalitz plot analysis of the final state is performed after integrating over the angular variable $\chi$.

\begin{figure}[t]
\centering
\includegraphics[width=0.78\textwidth]{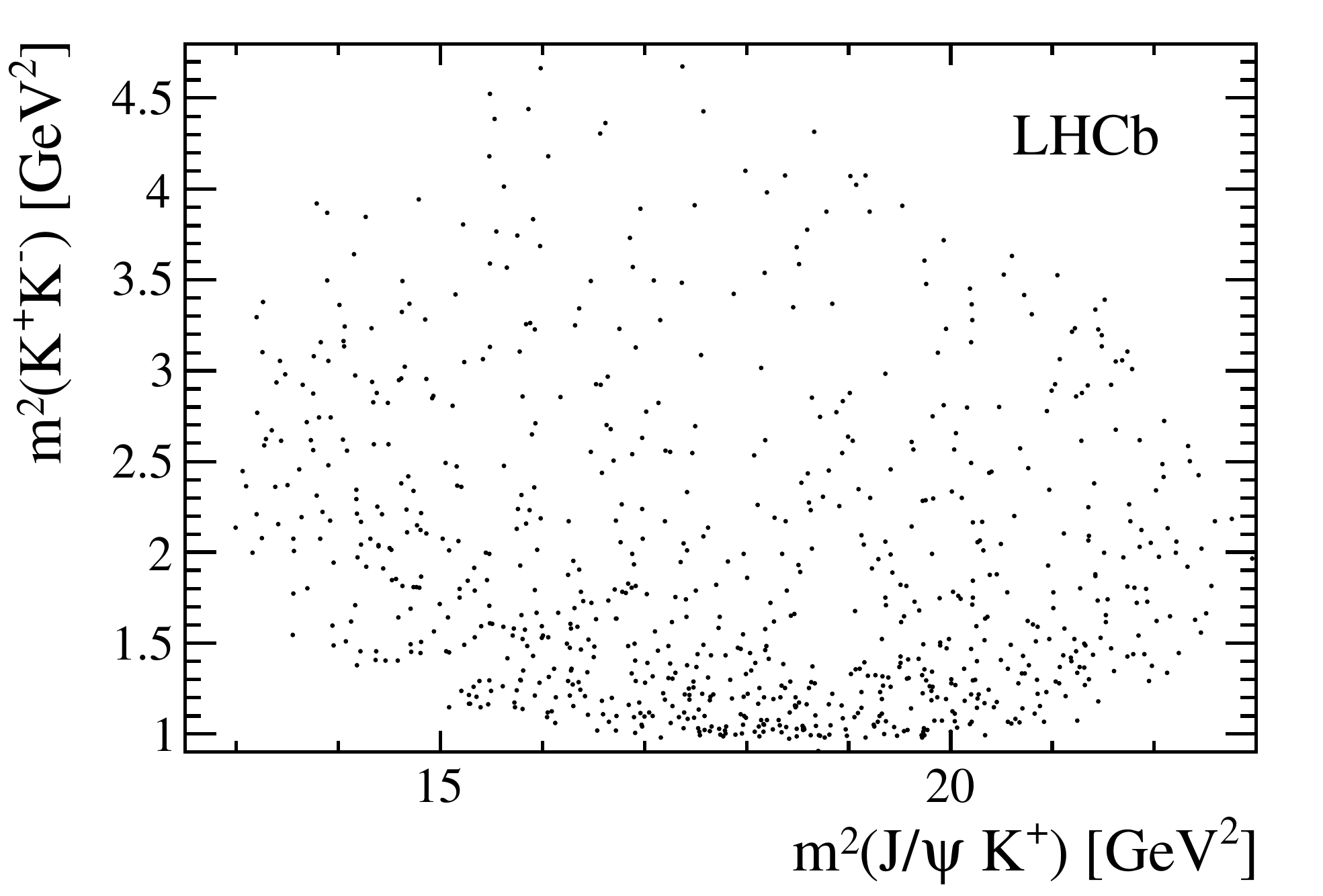}
\vskip -3mm
\caption{\small Distribution of $m^2(\KpKm)$ versus $m^2(\jpsi \Kp)$ for $\jpsi\KpKm$ candidates with mass within $\pm 20$ MeV of the $\Bdb$ mass.}
\label{fig:dalitz}
\end{figure}

To study the resonant structures of the decay $\Bdb\to \jpsi \KpKm$, we use  candidates  with  invariant mass within $\pm 20$ MeV of the observed $\Bdb$ mass peak.
The invariant mass squared of $\KpKm$ versus $\jpsi \Kp$   is shown in Fig.~\ref{fig:dalitz}. 
An excess of events is  visible at  low $\KpKm$ mass, which could include both nonresonant and resonant contributions.
Possible resonance candidates include $a_0(980)$, $f_0(980)$, $\phi$, $f_0(1370)$, $a_0(1450)$, or $f_0(1500)$ mesons.  Because of the limited sample size, we  perform the  analysis including only   the $a_0(980)$ and  $f_0(980)$ resonances  and  nonresonant components. 

In our previous analysis of $\Bdb\to\jpsi\pi^+\pi^-$ decay~\cite{Aaij:2013zpt}, we did not see a statistically significant contribution  of the $f_0(980)$ resonance. The branching fraction product was determined as
\begin{equation*}
\mathcal{B}(\Bdb\to\jpsi f_0(980),~f_0(980)\to\pi^+\pi^-)=(6.1~^{+3.1}_{-2.0}~      ^{+1.7}_{-1.4})\times 10^{-7}.
\end{equation*}
Using this branching fraction product and  the  ratio of branching fractions, 
\begin{equation}
\mathcal{R} = \frac{\mathcal{B}(f_0(980)\to \KpKm)}{\mathcal{B}(f_0(980)\to \pip\pim)}=0.35^{+0.15}_{-0.14},
\end{equation}
determined from an average of the BES~\cite{Ablikim:2005kp} and \babar~\cite{Aubert:2006nu} measurements, we  estimate the expected yield of  $\Bdb\to \jpsi f_0(980)$ with $f_0(980)\to \KpKm$ as
\begin{equation*}
N(\Bdb\to \jpsi f_0(980),~ f_0(980)\to \KpKm) = 20^{+14}_{-11}.
\end{equation*}

Although the $f_0(980)$ meson is easier to detect in its $\pip\pim$ final state than in $\KpKm$, the presence of the $f_0(980)$ resonance was not established in the   $\Bdb\to\jpsi\pi^+\pi^-$ decay~\cite{Aaij:2013zpt}, despite   some positive indication. Therefore, we test for two models: one that includes  the $f_0(980)$ resonance  with fixed amplitude strength corresponding to the expected yield and label it as  ``default'' and the other without the $f_0(980)$ resonance. The latter is called ``alternate''.

\subsection{The model for $\Bdb\to \jpsi K^+K^-$ }
The overall probability density function (PDF) given by the sum of signal, $S$, and background functions, $B$, is 
\begin{eqnarray}
\label{eq:pdf}
F(s_{12}, s_{23}, \theta_{\jpsi})&=&\frac{1-f_{\rm com }-f_{\rm refl}}{{\cal{N}}_{\rm sig}}\varepsilon(s_{12}, s_{23}, \theta_{\jpsi}) S(s_{12}, s_{23}, \theta_{\jpsi})\\ 
&+& B(s_{12}, s_{23}, \theta_{\jpsi})
,\nonumber 
\end{eqnarray}
where the background is the sum of combinatorial background, $C$, and reflection, $R$, functions,
\begin{equation}
\label{eq:background}
B(s_{12}, s_{23}, \theta_{\jpsi})= \frac{f_{\rm com}}{{\cal{N}}_{\rm com}} C(s_{12}, s_{23},  \theta_{\jpsi})+ \frac{f_{\rm refl}}{{\cal{N}}_{\rm refl}} R(s_{12}, s_{23}, \theta_{\jpsi}),
\end{equation}
and $f_{\rm com}$ and $f_{\rm refl}$ are the fractions of the combinatorial background and reflection, respectively, in the fitted region, and $\varepsilon$ is the detection efficiency. The fractions $f_{\rm com}$ and $f_{\rm refl}$, obtained from the mass fit, are fixed for the subsequent analysis.

The normalization factors are given by
\begin{eqnarray}
{\cal{N}}_{\rm sig}&=&\int \! \varepsilon(s_{12}, s_{23}, \theta_{\jpsi}) S(s_{12}, s_{23}, \theta_{\jpsi}) \, 
ds_{12}ds_{23}d \cos \theta_{\jpsi},\nonumber\\
{\cal{N}}_{\rm com}&=&\int \!C(s_{12}, s_{23}, \theta_{\jpsi}) \, 
ds_{12}ds_{23}d\cos \theta_{\jpsi},\\
{\cal{N}}_{\rm refl}&=&\int \!R(s_{12}, s_{23}, \theta_{\jpsi}) \, 
ds_{12}ds_{23}d\cos \theta_{\jpsi}\nonumber.
\end{eqnarray}

The expression for the signal function, $S(s_{12},s_{23},\theta_{\jpsi})$, amplitude for the nonresonant process  and other details of the fitting procedure are the same as used in the analysis described in Refs.~\cite{:2012cy,Aaij:2013orb,Aaij:2013zpt}. The amplitudes for the  $a_0(980)$ and $f_0(980)$ resonances are described below.

The main decay channels of the $a_0(980)$ (or $f_0(980)$)  resonance are $\eta \pi$ (or $\pi \pi$) and $K\Kb$,  with the former being the larger~\cite{PDG}. Both the $a_0(980)$ and the $f_0(980)$ resonances are very close to the $K\Kb$ threshold, which can strongly influence the resonance shape.  To take this complication into account, we follow the  widely accepted prescription proposed by Flatt\'e~\cite{Flatte:1976xv}, based on the coupled channels $\eta \pi^0$ (or $\pi \pi$) and $KK$. The Flatt\'e mass shapes are parametrized as
\begin{equation}
A^{a_0}_R(s_{23})= \frac{1}{m_R^2-s_{23}-i(g_{\eta\pi}^2\rho_{\eta\pi}+g_{KK}^2\rho_{KK})} 
\end{equation}
for the $a_0(980)$ resonance, and 
\begin{equation}
A^{f_0}_R(s_{23})= \frac{1}{m_R^2-s_{23}-im_R(g_{\pi\pi}\rho_{\pi\pi}+g_{KK}\rho_{KK})}
\end{equation}
for the $f_0(980)$ resonance. In both cases, $m_R$  refers to the pole mass of the resonance. The constants $g_{\eta\pi}$ (or $g_{\pi\pi}$) and $g_{KK}$ are the  coupling strengths of $a_0(980)$ (or $f_0(980)$) to $\eta\pi^0$ (or $\pi\pi$) and $KK$ final states, respectively. The $\rho$ factors are given by the Lorentz-invariant phase space
\begin{eqnarray}
\rho_{\eta\pi} &=& \sqrt{\left(1-\left(\frac{m_{\eta}-m_{\pi^0}}{\sqrt{s_{23}}}\right)^2\right)\left(1-\left(\frac{m_{\eta}+m_{\pi^0}}{\sqrt{s_{23}}}\right)^2\right)}\label{flatte1}, \\ 
\rho_{\pi\pi} &=& \frac{2}{3}\sqrt{1-\frac{4m^2_{\pi^{\pm}}}{s_{23}}}+\frac{1}{3}\sqrt{1-\frac{4m^2_{\pi^{0}}}{s_{23}}},\\
\rho_{KK} &=& \frac{1}{2}\sqrt{1-\frac{4m^2_{K^{\pm}}}{s_{23}}}+\frac{1}{2}\sqrt{1-\frac{4m^2_{K^{0}}}{s_{23}}}.\label{flatte2}
\end{eqnarray}

The  parameters for the $a_0(980)$ lineshape are fixed in the fit as  determined by the Crystal Barrel experiment~\cite{Abele:1998qd}. The parameters are $m_R=999\pm 2~\rm MeV$, $g_{\eta\pi}=324\pm 15~\rm MeV$ and $g^2_{KK}/g^2_{\eta\pi}=1.03\pm 0.14$. The  parameters for $f_0(980)$  are also fixed to the values $m_R=939.9\pm6.3$ MeV, $g_{\pi\pi}=199\pm30$ MeV and $g_{KK}/g_{\pi\pi}=3.0\pm0.3$, obtained from our previous analysis of $\Bsb\to\jpsi\pip\pim$ decay~\cite{:2012cy}.
\subsection{Detection efficiency}
\label{sec:mc}
The detection efficiency is determined from a sample of $10^{6}$ $\Bdb\rightarrow \jpsi\KpKm$ simulated  events  that are generated uniformly in phase space. The distributions of the generated \Bdb meson are weighted according to the $p$ and $\pt$  distributions in order to  match those observed in data. We also  correct  for the differences between the simulated kaon detection efficiencies and the measured ones determined by using a sample of  $D^{*+}\to \pip (D^0\to K^-\pip)$ events. 

The efficiency is  described in terms of the analysis variables. 
Both $s_{12}$ and $s_{13}$  range from  $12.5~\rm GeV^2$  to $23.0~\rm GeV^2$, where $s_{13}$ is defined below,  and thus are centered at $s_0=17.75$ GeV$^2$.
We model the detection efficiency using the dimensionless symmetric Dalitz plot observables
\begin{equation}
x= (s_{12}-s_0)/(1~{\rm GeV}^2)~~~~{\rm and}~~~~  y=(s_{13}-s_0)/(1~{\rm GeV}^2),\label{eq:eff_variable}
\end{equation}
and the angular variable $\theta_{\jpsi}$. The observables $s_{12}$ and $s_{13}$  are related to $s_{23}$ as
\begin{equation}
s_{12}+s_{13}+s_{23}=m^2_B+m^2_{\jpsi}+m^2_{K^+}+m^2_{K^-}~.\label{conver}
\end{equation}
\begin{figure}[t]
\centering
\includegraphics[width=0.78\textwidth]{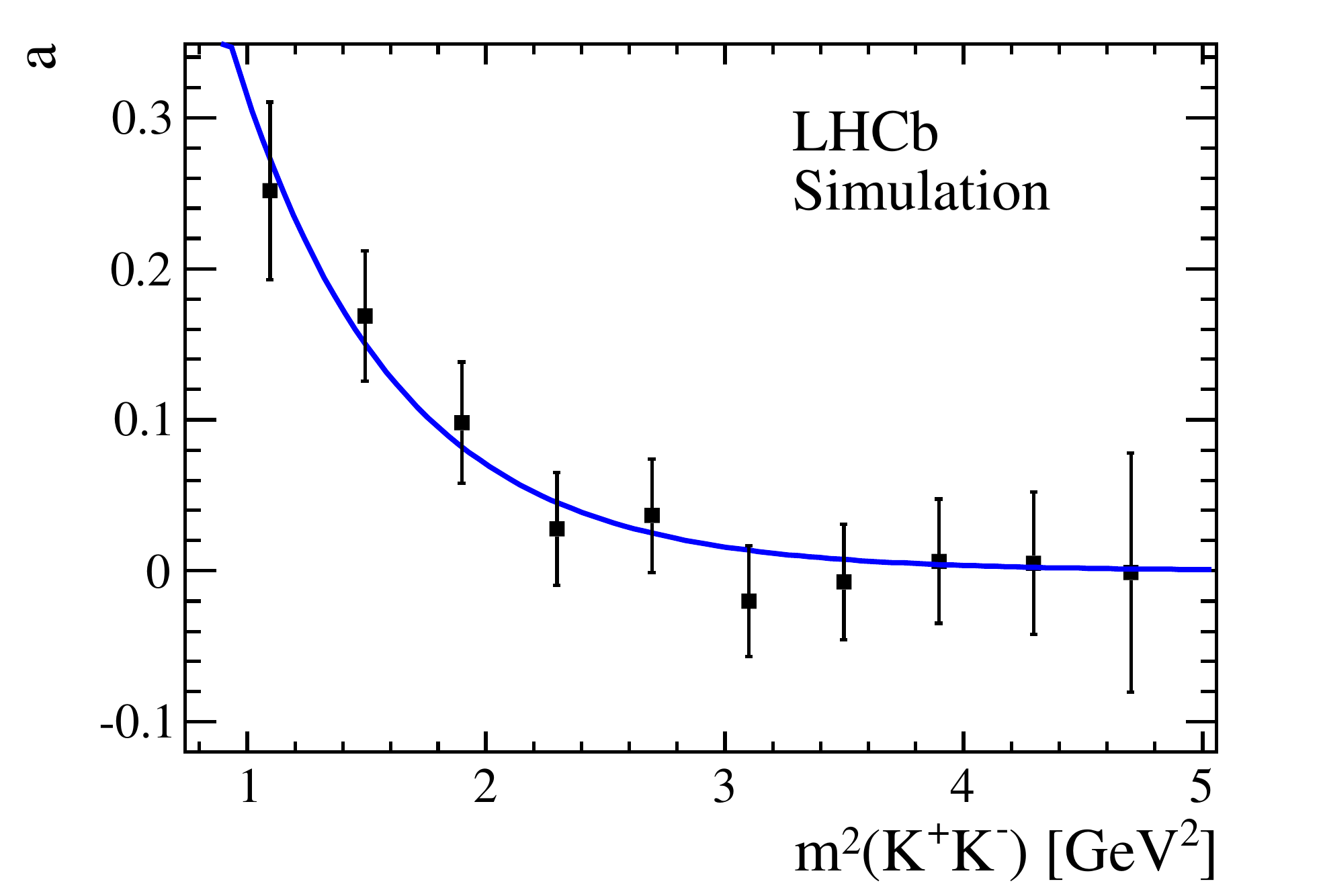}
\vskip -3mm
\caption{\small Exponential fit to the acceptance parameter $a(s_{23})$.}
\label{fig:cosHacc}
\end{figure}

To parametrize this efficiency, we fit the $\cos \theta_{\jpsi}$ distributions of the $\Bdb \to  \jpsi \KpKm$ simulated sample in bins of $s_{23}$ with the function
\begin{equation}
\varepsilon_2(s_{23},\theta_{\jpsi})=\frac{1+ a(s_{23})\cos^2\theta_{\jpsi}}{2+2a(s_{23})/3},\label{eq:cosHacc}
\end{equation}
where $a$ is a function of $s_{23}$. The resulting distribution, shown in Fig.~\ref{fig:cosHacc},  is  described by an exponential function

\begin{equation}
a(s_{23})= \exp(a_1+a_2 s_{23}),
\end{equation}
where $a_1$ and $a_2$ are  constant parameters.
Equation~(\ref{eq:cosHacc}) is normalized to one when integrated over $\cos \theta_{\jpsi}$. The efficiency as a function of  $\cos \theta_{\jpsi}$ also depends on $s_{23}$, and is observed to be independent of $s_{12}$. 
Therefore, the  detection efficiency  can be expressed as
\begin{equation}
\varepsilon(s_{12}, s_{23}, \theta_{\jpsi})=\varepsilon_1(x, y)\times \varepsilon_2(s_{23}, \theta_{\jpsi}).\label{eq:eff}
\end{equation}
After integrating over $\cos \theta_{\jpsi}$,  Eq.~(\ref{eq:eff}) becomes
\begin{equation}
\int_{-1}^{+1}\varepsilon(s_{12}, s_{23}, \theta_{\jpsi})d\cos \theta_{\jpsi}=\varepsilon_1(x, y).
\end{equation}
and is modeled by a  symmetric fourth-order polynomial function given by
\begin{eqnarray}
\varepsilon_1(x, y)&=& 1+\epsilon'_1(x+y)+\epsilon'_2(x+y)^2+\epsilon'_3xy+\epsilon'_4(x+y)^3
+\epsilon'_5 xy(x+y)\nonumber \\
&&+\epsilon'_6(x+y)^4+\epsilon'_7 xy(x+y)^2+\epsilon'_8 x^2y^2,
\end{eqnarray}
where the $\epsilon'_i$ are  fit parameters.
\begin{figure}[t]
\begin{center}
     \includegraphics[width=0.78\textwidth]{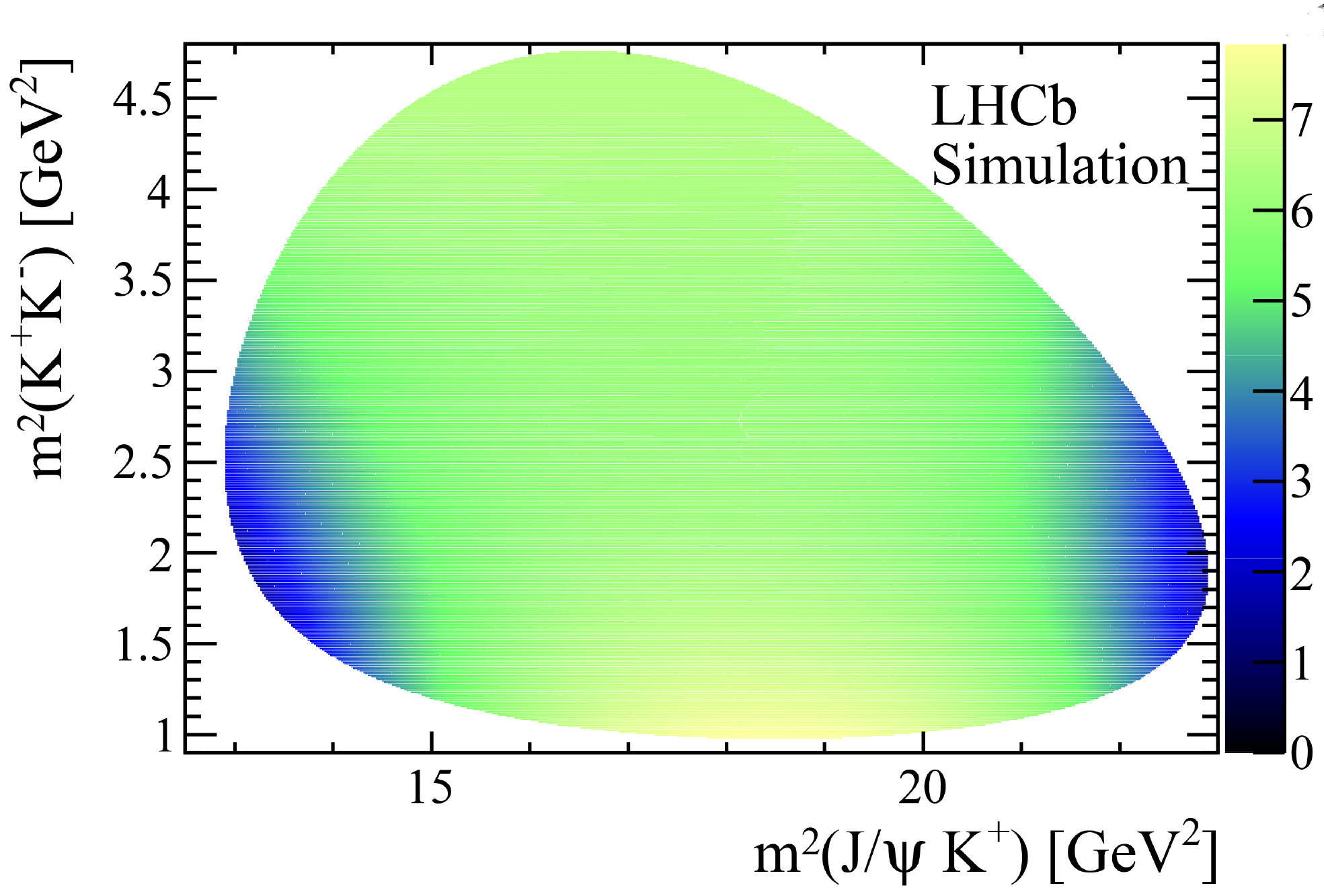}
\end{center}\label{eff1}
\vskip -3mm
\caption{\small Parametrized detection efficiency as a function
  of  $m^2(\KpKm)$ versus $ m^2(\jpsi\Kp)$. The $z$-axis scale is arbitrary.}
\end{figure}
\begin{figure}[h]
\begin{center}
    \includegraphics[width=0.48\textwidth]{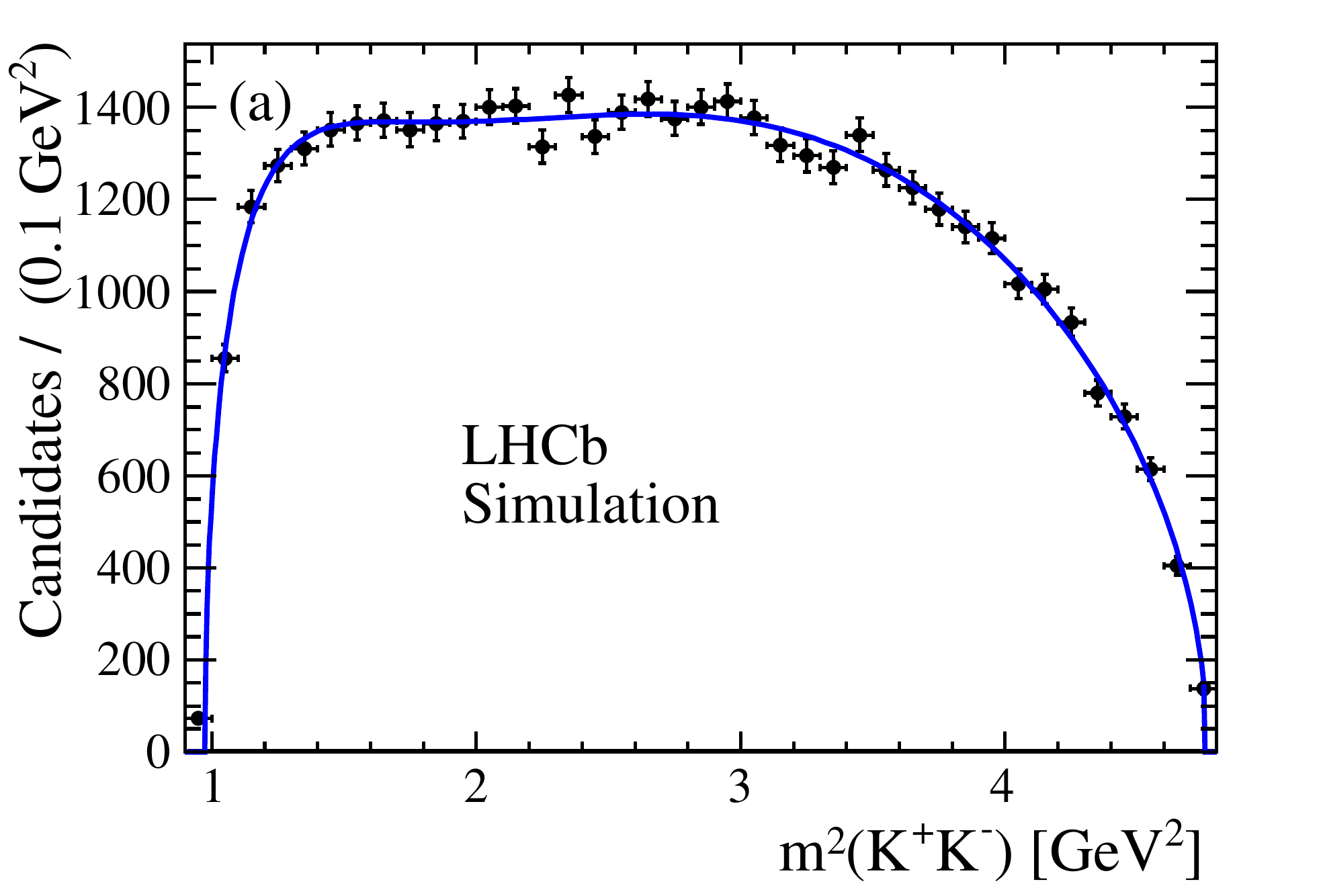}%
    \includegraphics[width =0.48\textwidth]{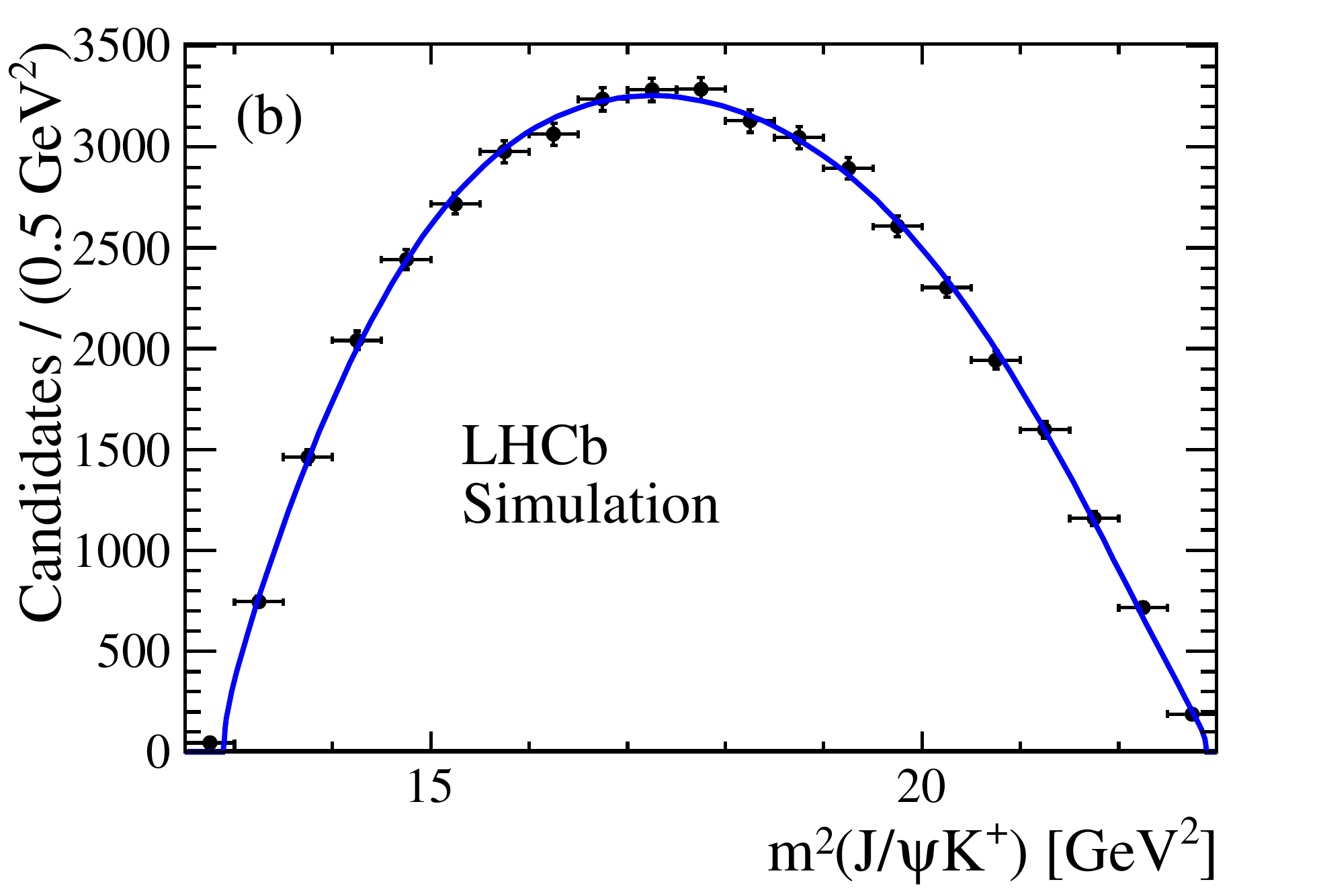}
\end{center}\label{eff2}
\vskip -0.5cm
\caption{\small Projections  of (a) $m^2(\KpKm)$ and (b)  $ m^2(\jpsi \Kp)$ of the simulated Dalitz plot used to measure the efficiency parameters. The points represent the simulated event distributions, and the curves the projections of the polynomial fit.}
\end{figure}

Figure~\ref{eff1} shows the polynomial function obtained from a fit to the Dalitz plot distributions of simulated events.
The projections of the fit  describe  the efficiency well as can be seen  in Fig.~\ref{eff2}. 
\subsection{Background composition}
\label{sec:background}
To parametrize the combinatorial background, we use the $\Bdb$ mass sidebands, defined as the regions from 35 MeV to 60 MeV on the lower side and 25 MeV to 40 MeV on the upper side of the $\Bdb$ mass peak.
The  shape of the combinatorial background is found to be
\begin{equation}
C(s_{12}, s_{23}, \theta_{\jpsi})=\left[C_1(s_{12}, s_{23})\frac{P_B}{m_B}+\frac{c_0}{(m^2_0-s_{23})^2+m^2_0\Gamma_0^2}\right]\times \left(1+\alpha \cos^2 \theta_{\jpsi}\right),
\end{equation}
with $C_1(s_{12}, s_{23})$   parametrized as
\begin{equation}
C_1(s_{12}, s_{23}) = 1+c_1(x+y)+c_2(x+y)^2+c_3xy+c_4(x+y)^3
+c_5xy(x+y),
\end{equation}
where $P_B$ is the magnitude of the $\jpsi$ three-momentum in the $\Bdb$ rest frame, $m_B$ is the known $\Bdb$ mass, and $c_i$, $m_0$, $\Gamma_0$ and $\alpha$ are the model parameters.
The variables $x$ and $y$ are defined in Eq.~(\ref{eq:eff_variable}).

\begin{figure}[h]
\centering
\includegraphics[width=0.48\textwidth]{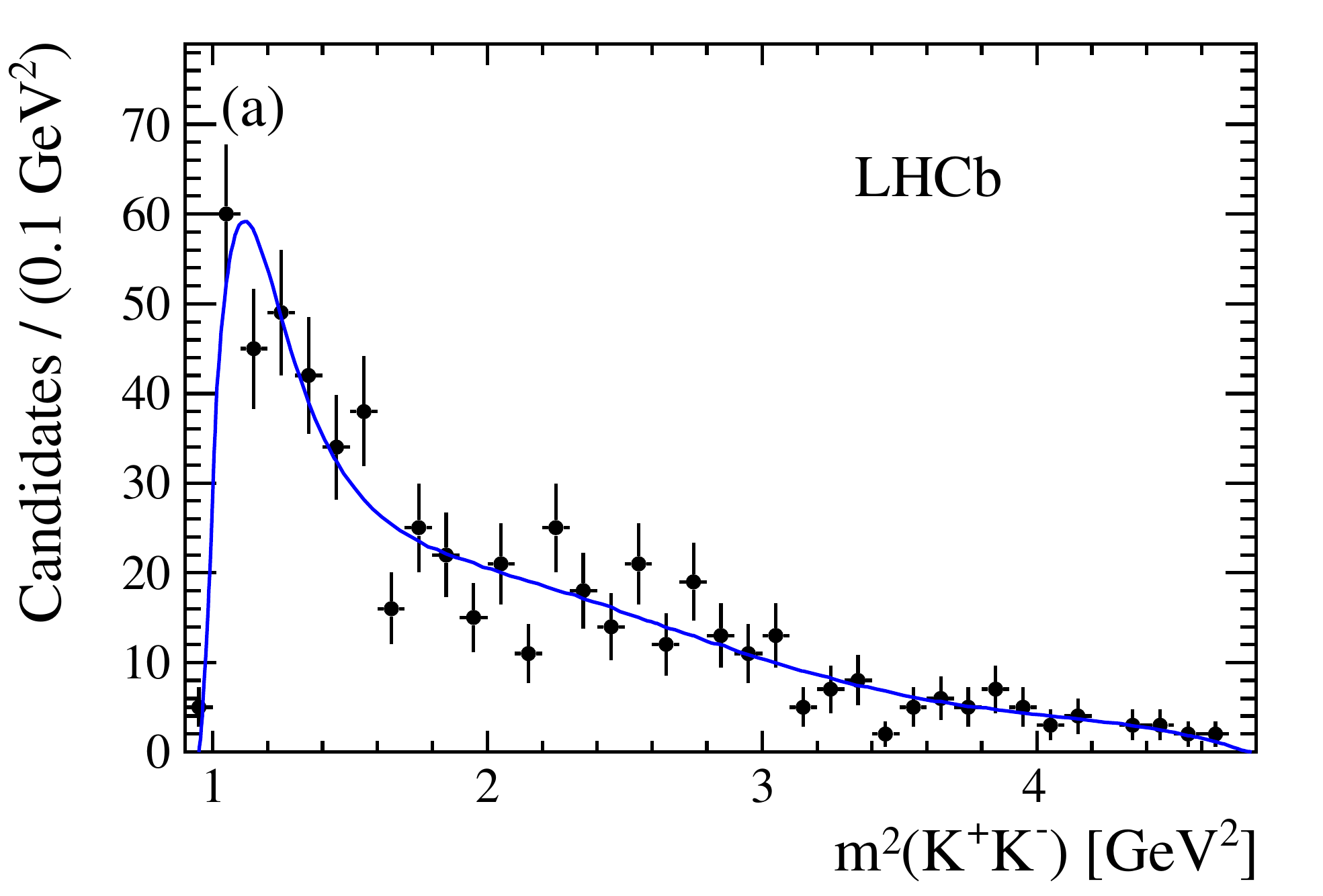}
\includegraphics[width=0.48\textwidth]{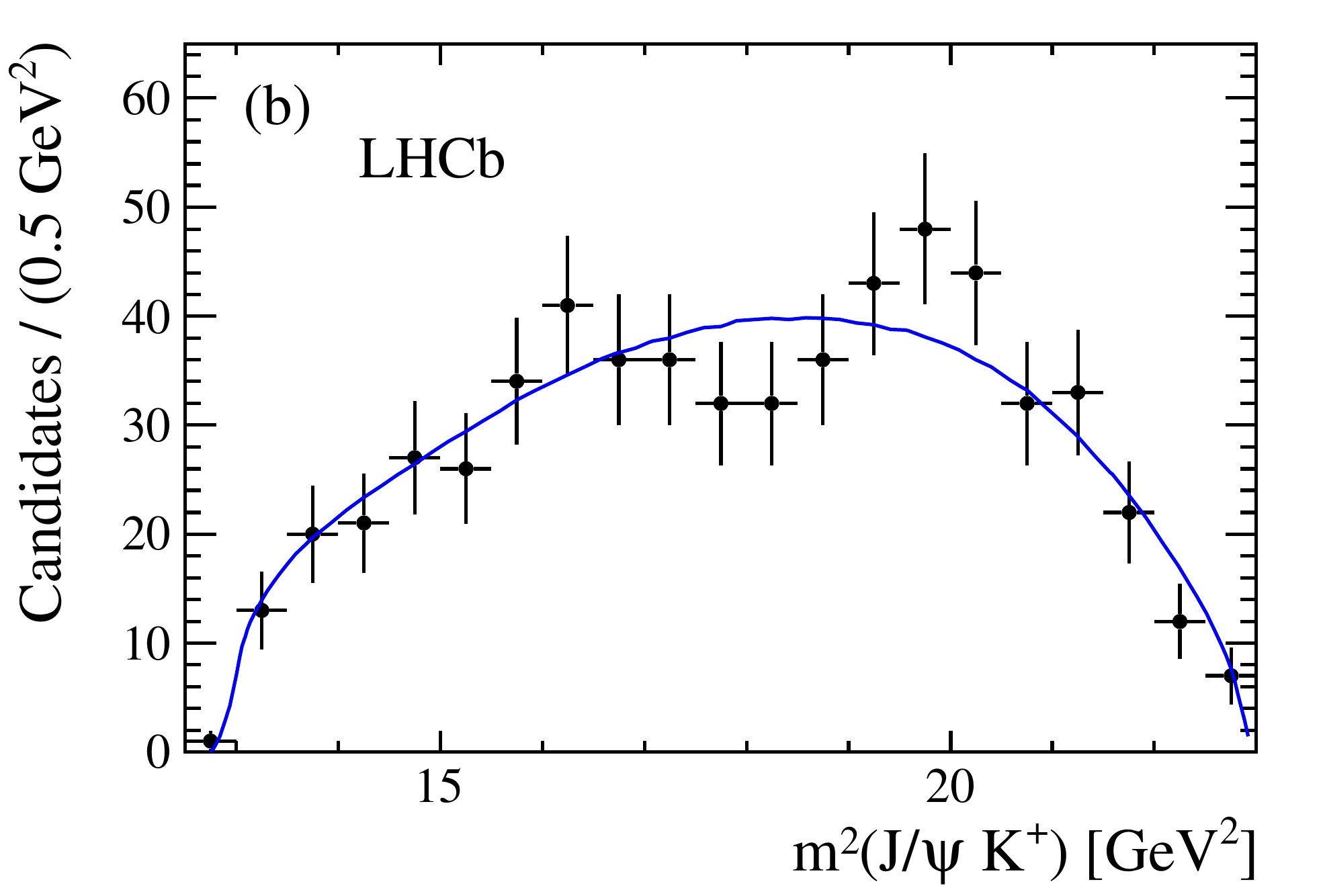}
\vskip -3mm
\caption{\small Invariant mass squared projections of (a)  $K^+K^-$, and (b) $\jpsi K^{+}$ from the Dalitz plot of candidates in the \Bdb mass sidebands, with fit projection overlaid.}
\label{fig:bkgmodel}
\end{figure}

\begin{figure}[h]
\centering
\includegraphics[width=0.78\textwidth]{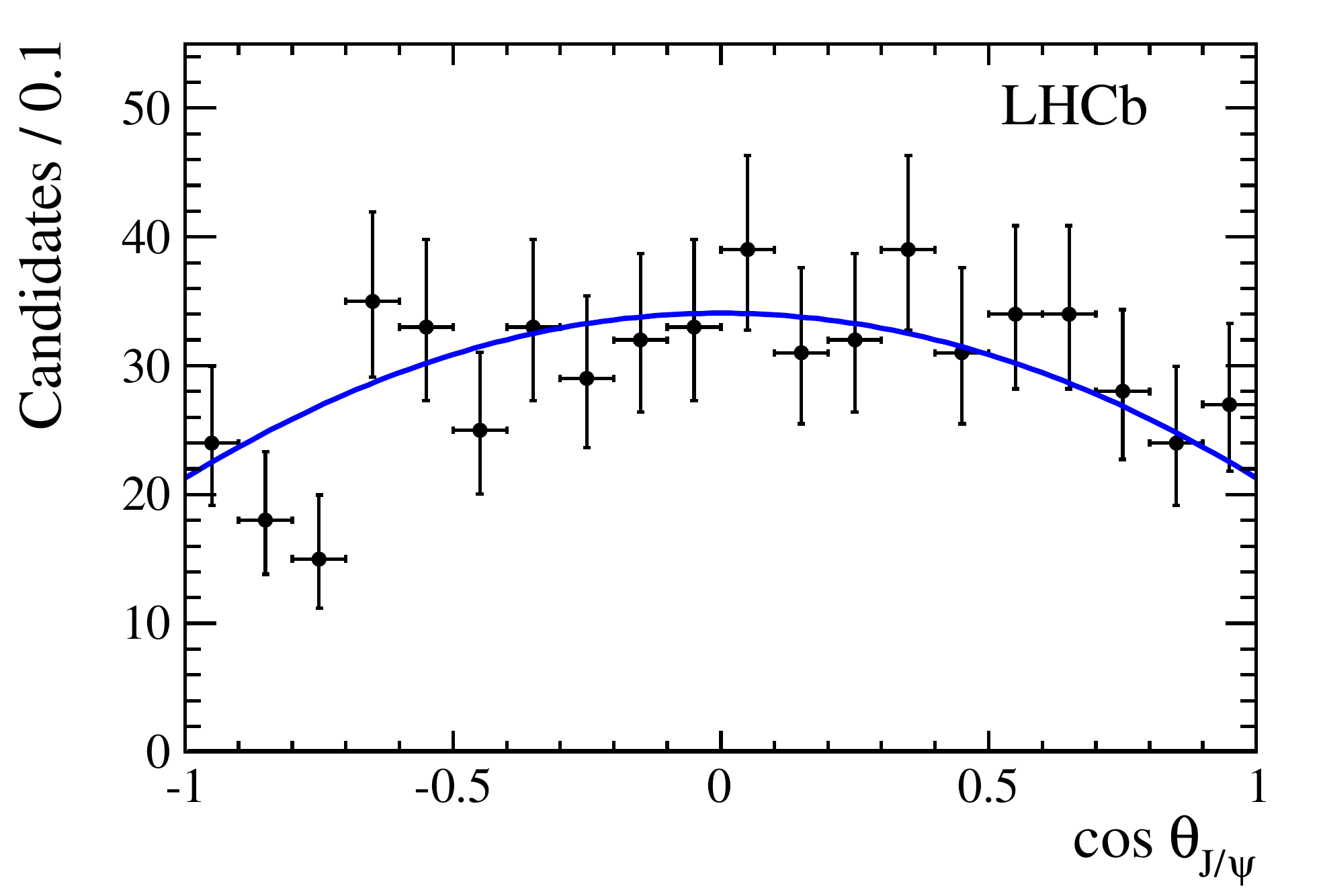}
\vskip -3mm
\caption{\small Distribution of   $\cos \theta_{\jpsi}$ from the  \Bdb mass sidebands, fitted with the function $1+\alpha \cos^{2} \theta_{\jpsi}$.}
\label{fig:bkgcosH}
\end{figure}

Figure~\ref{fig:bkgmodel} shows the invariant mass squared projections from an unbinned likelihood fit to the sidebands. 
The value of $\alpha$ is determined by fitting the $\cos\theta_{\jpsi}$ distribution of the combinatorial background sample, as shown in Fig. \ref{fig:bkgcosH}, with a function of the form $1+\alpha \cos^{2} \theta_{\jpsi}$, yielding $\alpha=-0.38\pm 0.10$.

\begin{figure}[h]
\centering
\includegraphics[width=0.48\textwidth]{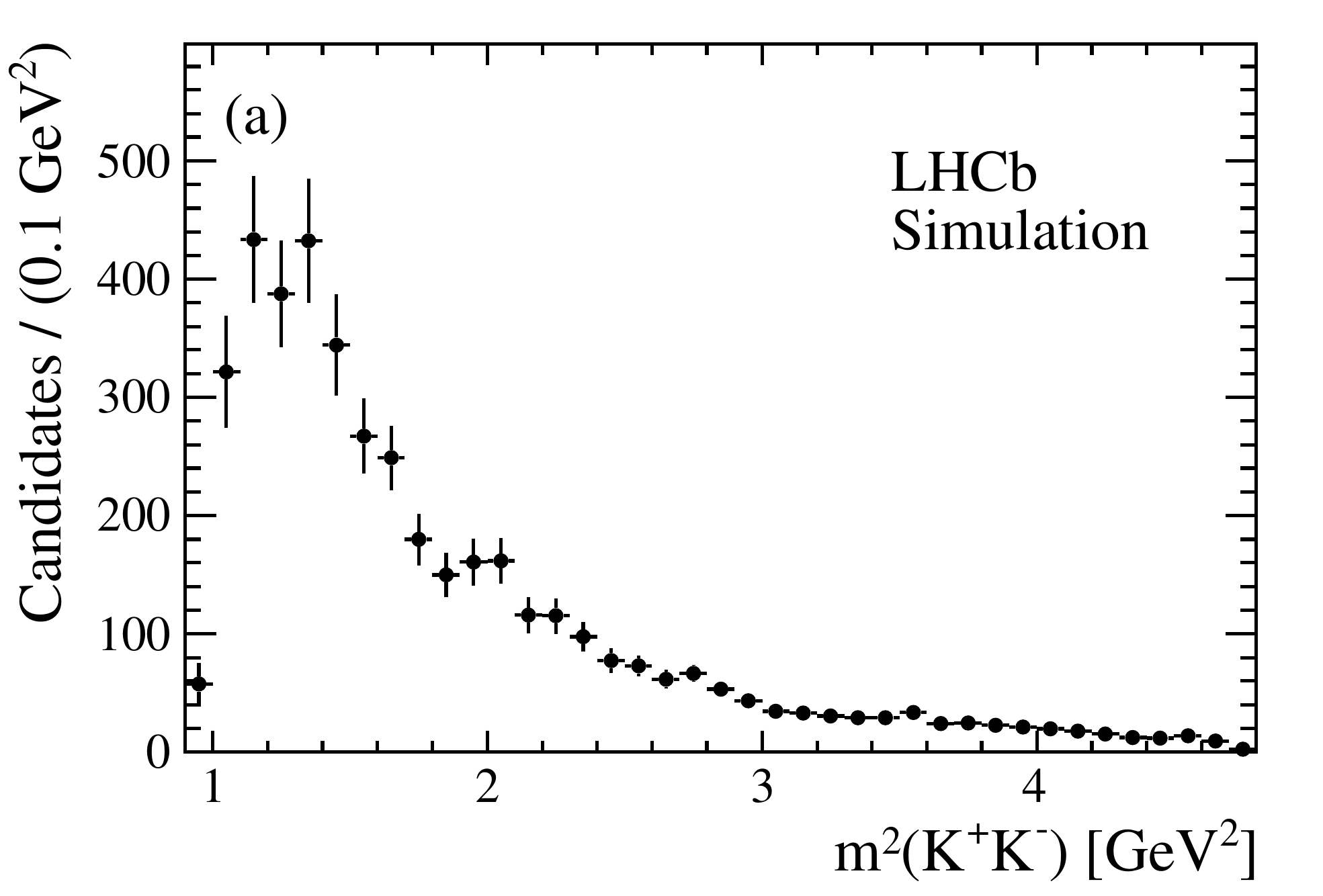}
\includegraphics[width=0.48\textwidth]{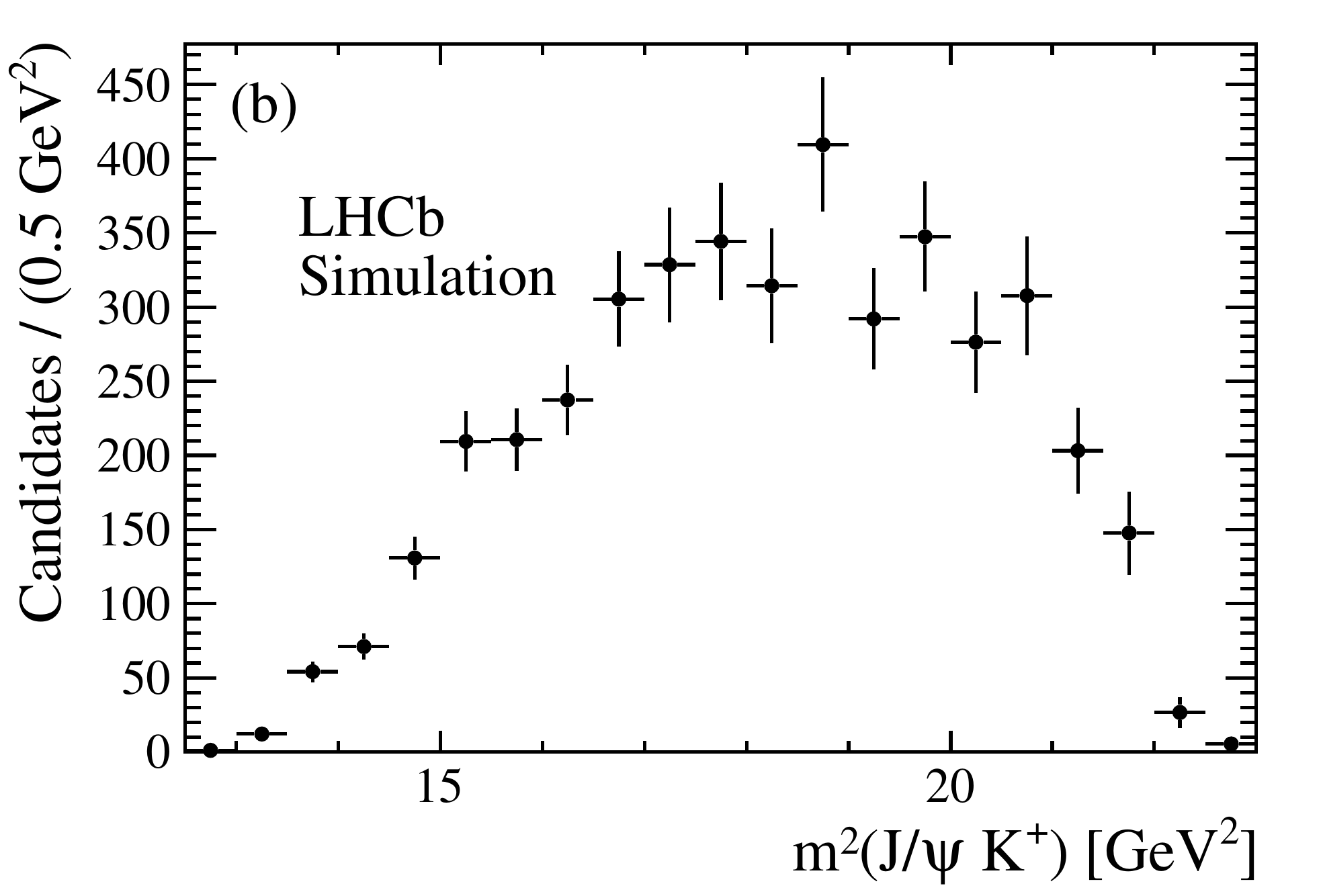}
\vskip -3mm
\caption{\small Projections of the reflection background in the variables (a) $ m^2(K^+K^-)$   and (b) $ m^2(\jpsi K^{+})$.}
\label{fig:refmodel}
\end{figure}

The reflection background is parametrized as
\begin{equation}
R(s_{12}, s_{23}, \theta_{\jpsi})=R_1(s_{12}, s_{23})\times  \left(1+\beta \cos^2 \theta_{\jpsi}\right), 
\end{equation}
where $R_1(s_{12}, s_{23})$ is modeled using the   simulation of $\Lb\to\jpsi p\Km$ decays weighted according to the $m(p\Km)$ distribution obtained in Ref.~\cite{Aaij:2013oha}. The projections are shown in Fig.~\ref{fig:refmodel}.
\begin{figure}[h]
\centering
\includegraphics[width=0.78\textwidth]{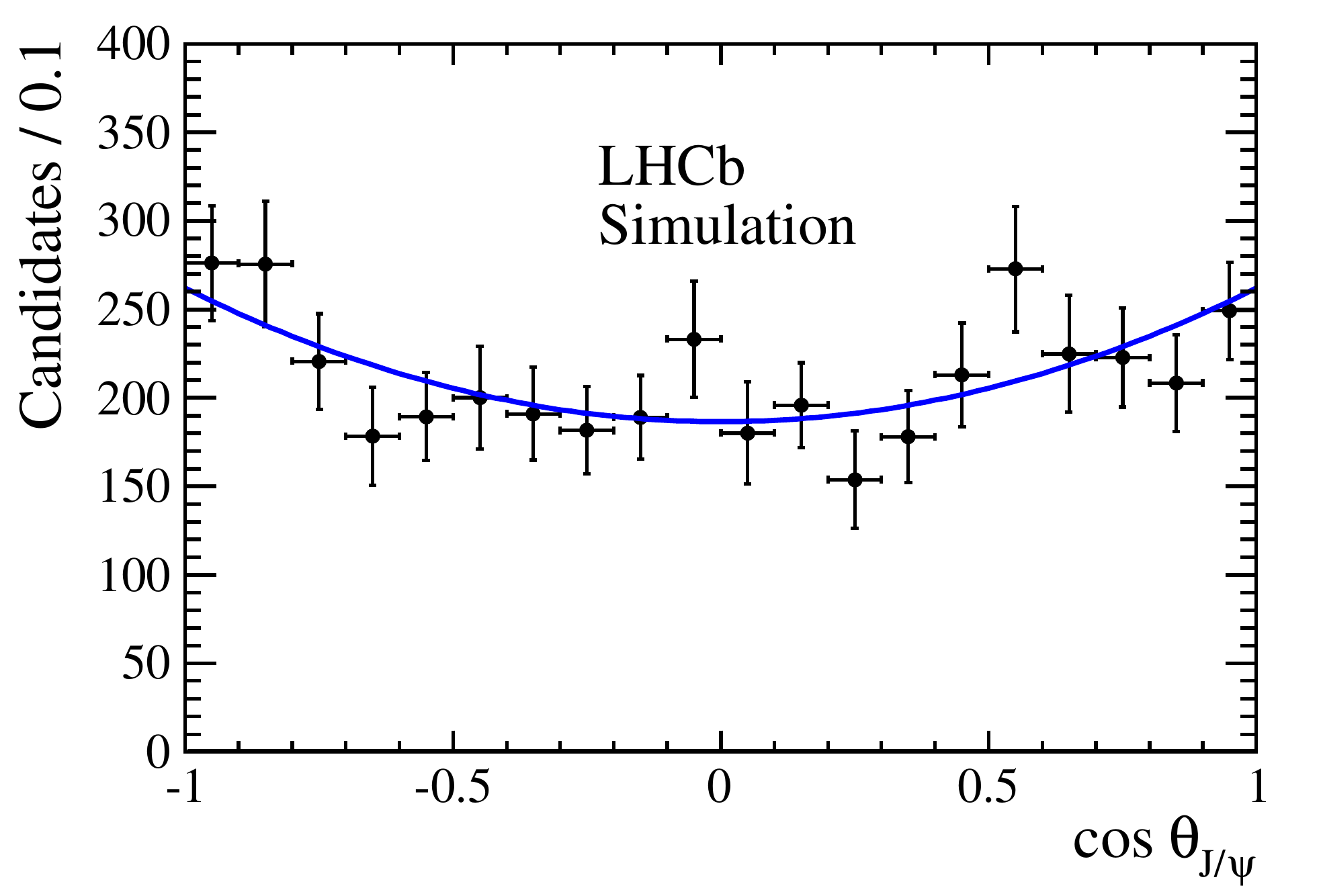}
\vskip -3mm
\caption{\small Distribution of  $\cos \theta_{\jpsi}$ for  the reflection background, fitted with the function $1+\beta \cos^{2} \theta_{\jpsi}$.}
\label{fig:refcosH}
\end{figure}
The $\jpsi$ helicity-dependent part of the reflection background parametrization is modeled as $1+\beta \cos^2 \theta_{\jpsi}$, where the parameter $\beta= 0.40\pm 0.08$ is obtained from a fit to the  simulated $\cos \theta_{\jpsi}$ distribution of the same sample, shown in Fig.~\ref{fig:refcosH}.

\subsection{Fit results}
\begin{figure}[t]
\centering
\includegraphics[width=0.78\textwidth]{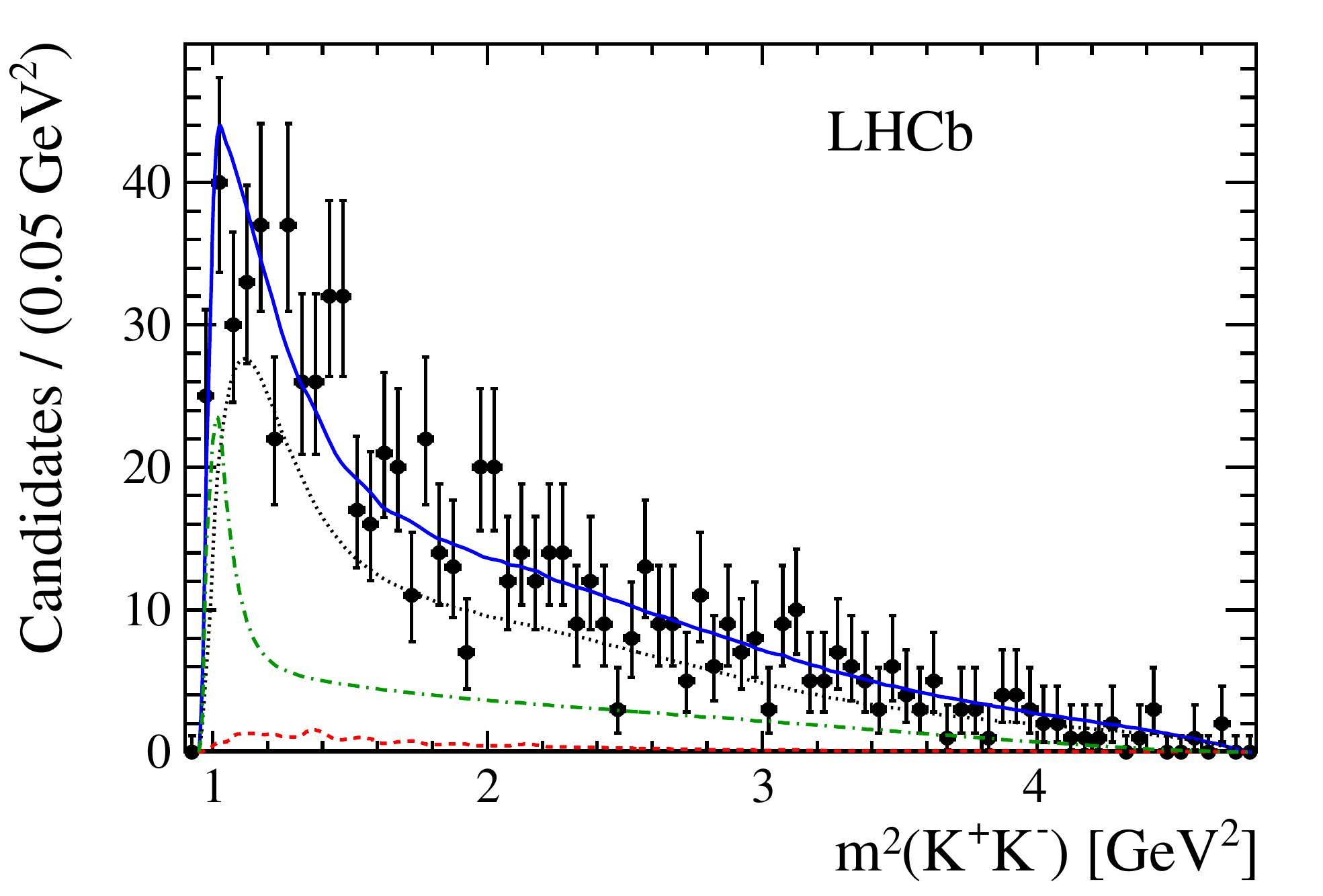}
\vskip -3mm
\caption{\small Dalitz plot fit projection of $m^2(\KpKm)$ in the signal region. The points with error bars are data, the (black) dotted curve shows the combinatorial background, the (red) dashed curve indicates the reflection from the misidentified $\Lb\to \jpsi p\Km$ decays, the (green) dot-dashed curve is the signal, and the (blue) solid line represents the total.}
\label{fig:modely}
\end{figure}

\begin{figure}[t]
\centering
\includegraphics[width=0.48\textwidth]{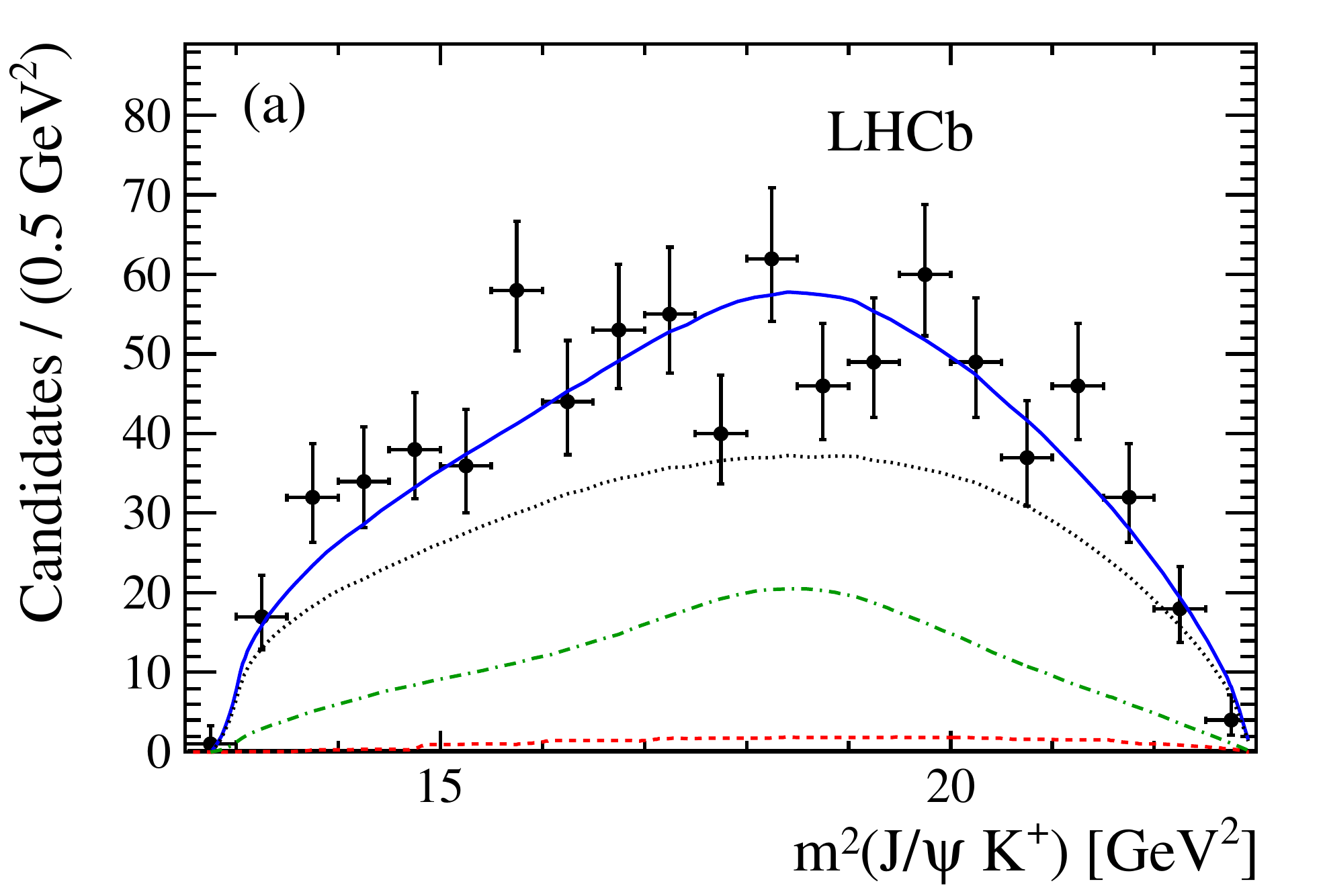}
\includegraphics[width=0.48\textwidth]{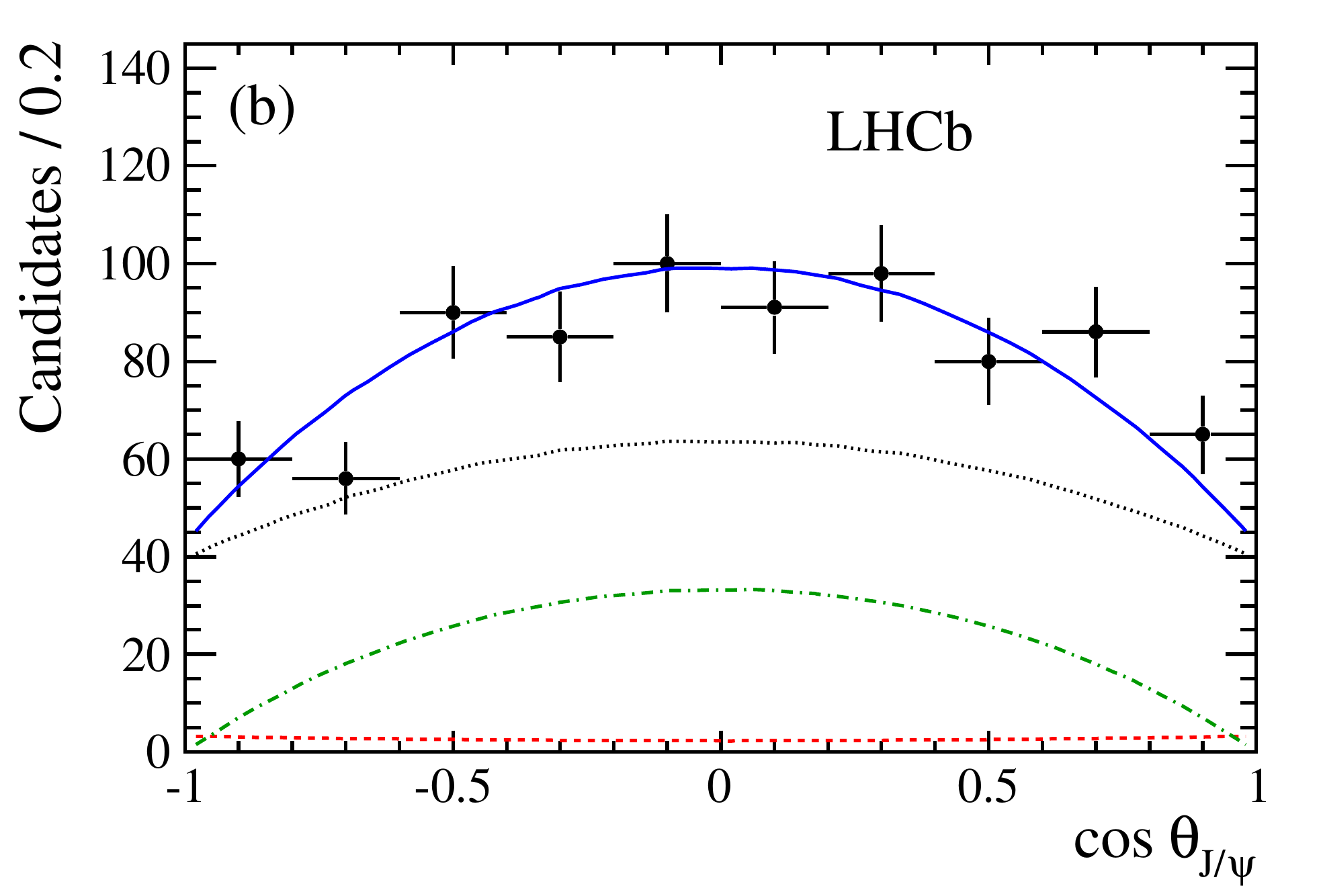}
\vskip -3mm
\caption{\small Dalitz plot fit projections of (a) $m^2(\jpsi K^{+})$ and (b) $\cos \theta_{\jpsi}$ in the signal region. The points with error bars are data, the (black) dotted curve shows the combinatorial background, the (red) dashed curve indicates the reflection from the misidentified $\Lb\to \jpsi p\Km$ decays, the (green) dot-dashed curve is the signal, and the (blue) solid line represents the total.}
\label{fig:modelxcosH}
\end{figure}

\begin{figure}[t]
\centering
\includegraphics[width=0.48\textwidth]{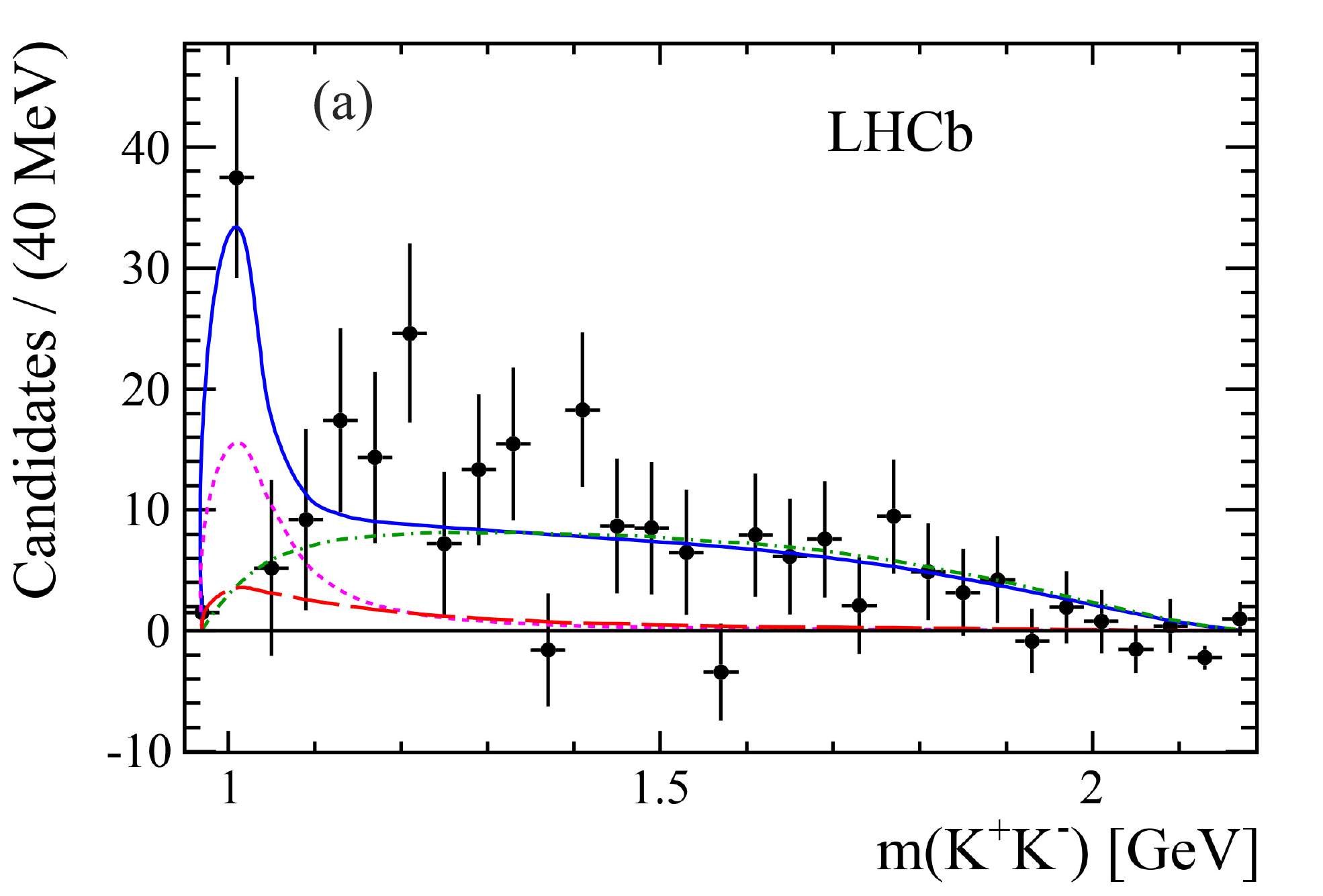}
\includegraphics[width=0.48\textwidth]{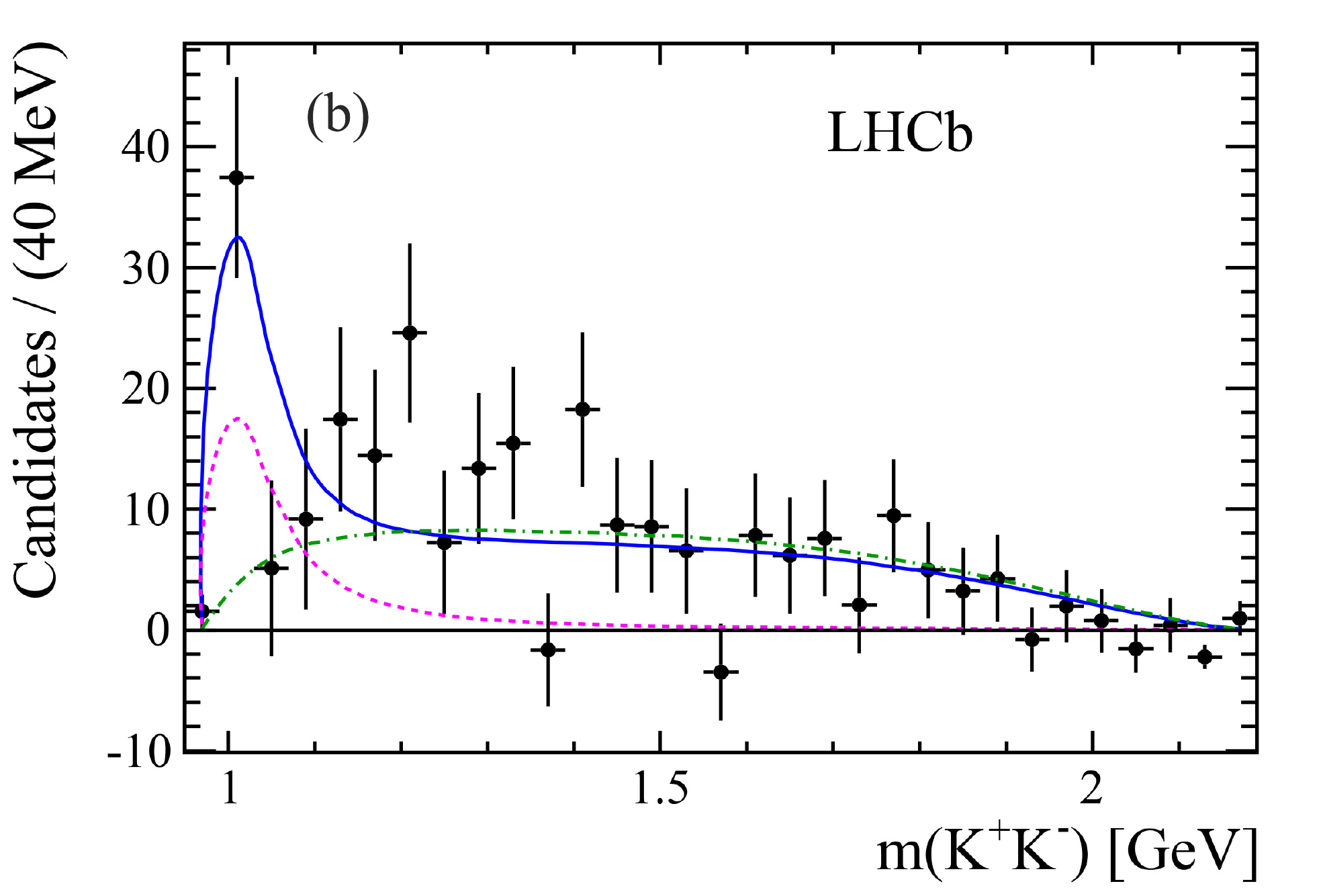}
\vskip -3mm
\caption{\small Background-subtracted $m(\KpKm)$ distributions for (a) default  and, (b) alternate fit models in the signal region.  The points with error bars are data, the (magenta) dashed curve shows the $a_0(980)$ resonance, the nonresonant contribution is shown by (green) dot-dashed curve and the (blue) solid curve represents the sum of $a_0(980)$, nonresonant and the interference between the two. The (red) long-dashed curve in (a) shows the $f_0(980)$ contribution.}
\label{fig:modelmkk}
\end{figure}

An unbinned maximum likelihood fit is performed  to extract the fit fractions and other physical parameters. 
Figure~\ref{fig:modely} shows the projection of $m^2(\KpKm)$ distribution for the default fit model. The $m^2(\jpsi \Kp)$ and the $\cos \theta_{\jpsi}$ projections are displayed in Fig.~\ref{fig:modelxcosH}. The background-subtracted $\KpKm$ invariant mass spectrum for default and alternate fit models  are shown in Fig.~\ref{fig:modelmkk}.  Both the combinatorial background and the reflection components of the fit are subtracted  from the data  to obtain the background-subtracted distribution. 

\begin{table}[h!t!p!]
\centering
\caption{\small Fit fractions  and phases   of the contributing components. The components of the form $\rm X+Y$ are the interference terms. Note that, in the default model the $f_0(980)$ amplitude strength is fixed to the expected value.  Poisson likelihood $\chi^2$~\cite{Baker:1983tu} is used to calculate the $\chi^2$.}
\vspace{0.2cm}
\begin{tabular}{ccccc}
\hline 
Component & ~~~~~~~~~~~~~Default &&  ~~~~~~~~~~~~~Alternate&\\
          & Fit fraction (\%) & Phase ($^\circ$) & Fit fraction (\%) & Phase ($^\circ$) \\
\hline
$a_0(980)$ & ~~$19\pm 13$&$-10\pm 27$  & ~$21\pm~8$& $-60\pm 26$\\
$f_0(980)$ & ~~$11\pm~5$& $-94\pm45$& ~-&~-\\
Nonresonant (NR) &  ~~$83\pm37$& 0 (fixed) &~$85\pm 23$& 0 (fixed)\\
 $a_0(980)$ + NR &   $-42\pm25$&~ - & $-~6\pm27$ &~ - \\
 $f_0(980)$ + NR & ~~$32\pm38$&~ -&~ -&~ -\\
 $a_0(980)$ + $f_0(980)$  & ~$-~2\pm16$&~ -&~ -&~ -\\
\hline
$- \rm ln \mathcal{L}$ & ~~~~~~~~~~~~~2940& & ~~~~~~~~~~~~~2943 &\\
$\chi^2/ \rm ndf$  &~~~~~~~~~~~~~1212/1406 &&~~~~~~~~~~~~~1218/1407& \\
\hline 
\end{tabular}
\label{tab:fitparameter}
\end{table}

The fit fractions and the phases of the contributing components for both models are given in Table~\ref{tab:fitparameter}. Quoted uncertainties are statistical only,  as determined from simulated experiments. 
We perform 500 experiments: each sample is generated according to the model PDF with input parameters  from the results of the  default fit. The correlations of the fitted parameters are also taken into account. For each experiment the fit fractions are calculated. The distributions of the obtained fit fractions are described by Gaussian functions. The r.m.s. widths of the Gaussian functions are taken as the statistical uncertainties on the corresponding parameters. 

The decay $\Bdb\to\jpsi \KpKm$  is dominated by  the nonresonant S-wave components in the $\KpKm$ system. The statistical significance of the $a_0(980)$ resonance is evaluated from the ratio,
$\mathcal{L}_{a_0+f_0+\rm NR}/\mathcal{L}_{f_0+\rm NR}$, of the maximum likelihoods obtained from the fits with and
without the resonance. The model with the resonance has two additional
degrees of freedoms, corresponding to the amplitude strength and the phase. The quantity $2{\rm ln}(\mathcal{L}_{a_0+f_0+\rm NR}/\mathcal{L}_{f_0+\rm NR})$
is found to be 18.6, corresponding to a significance of 3.9 Gaussian
standard deviations.  The large statistical uncertainty in the $a_0(980)$ fit fraction in the default model is due to the presence of the $f_0(980)$ resonance that is allowed to interfere with the $a_0(980)$ resonance whose phase is highly correlated with the fit fraction. This uncertainty is much reduced in the alternate model.

\begin{figure}[t]
\centering
    \includegraphics[width=0.48\textwidth]{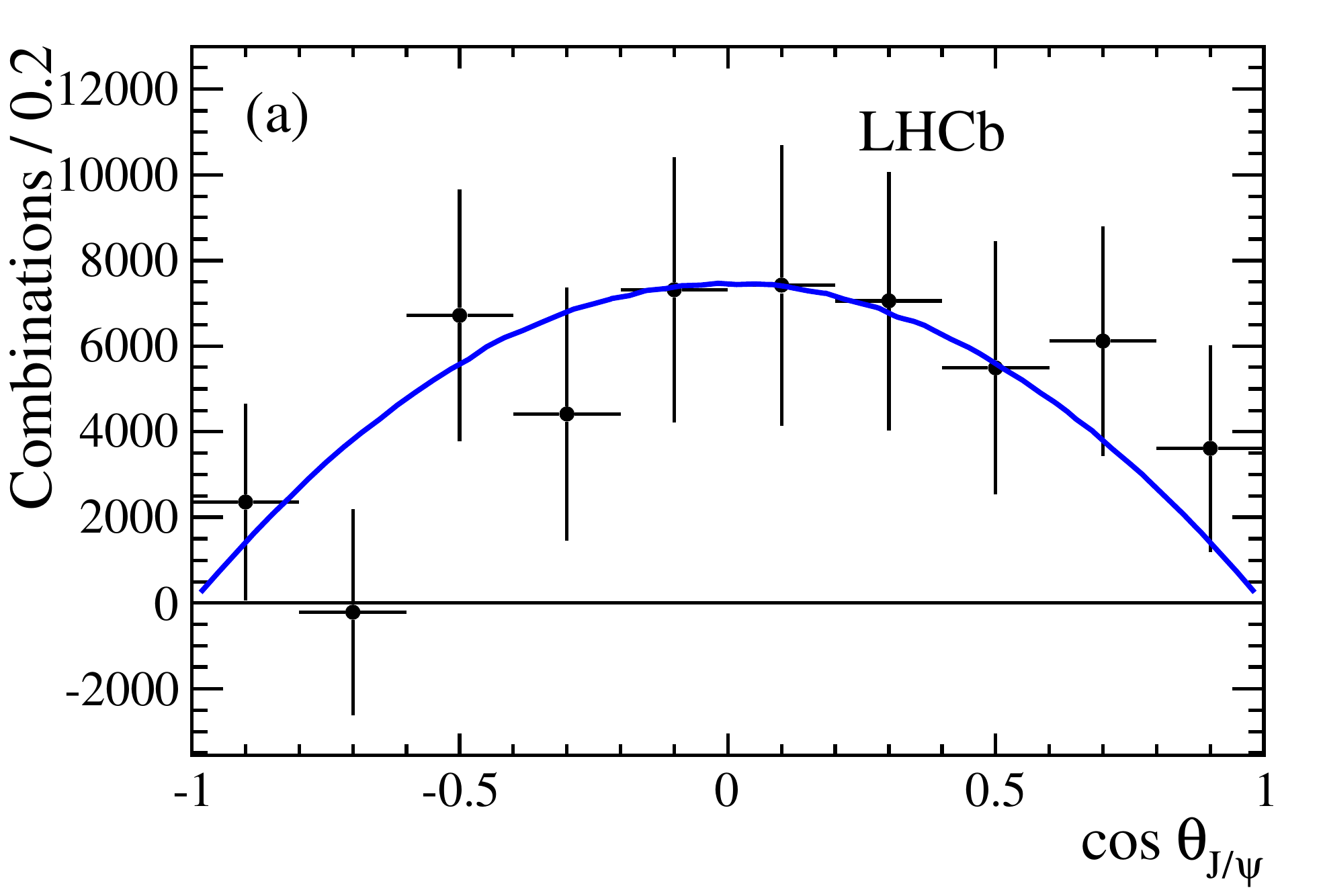}%
    \includegraphics[width=0.48\textwidth]{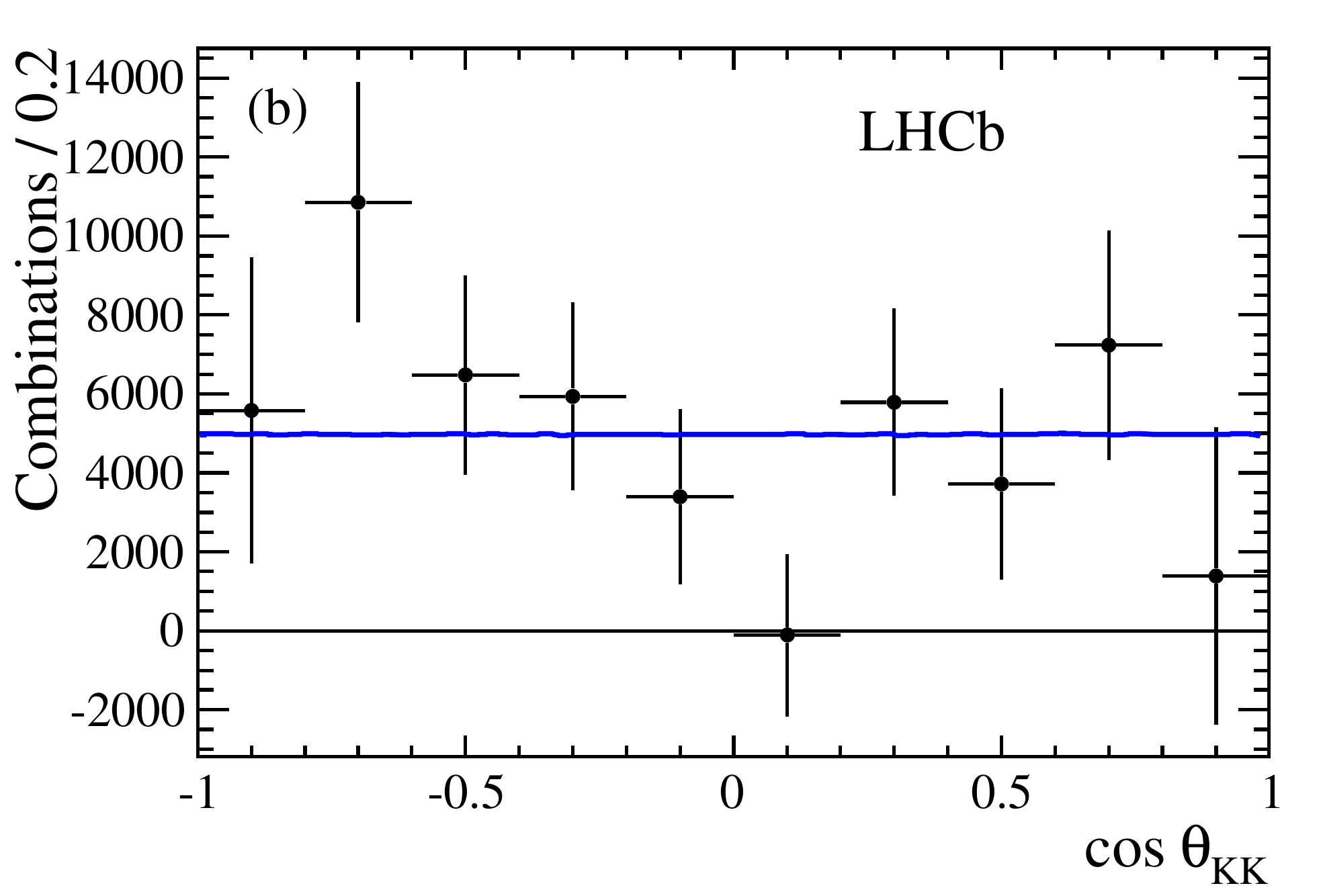}
\vskip -3mm
\caption{\small Background-subtracted and efficiency-corrected distribution of (a) $ \cos\theta_{\jpsi}$ ($\chi^2/\rm ndf=5.3/10$) and (b)  $\cos \theta_{KK}$ ($\chi^2/\rm ndf=12.8/10$). The points with error bars are data and the (blue) solid lines show the fit to the default model.}
\label{fig:helicity}
\end{figure}

The background-subtracted and efficiency-corrected  distributions of $ \cos\theta_{\jpsi}$ and $\cos \theta_{KK}$ are shown in Fig.~\ref{fig:helicity}. Since all the contributing components are  S-waves, the data should be distributed as $1-\cos^2\theta_{\jpsi}$ in $\cos\theta_{\jpsi}$ and uniformly in $\cos\theta_{KK}$. The $\cos \theta_{\jpsi}$ distribution follows the  expectation very well with $\chi^2/\rm ndf = 5.3/10$ and the $\cos \theta_{KK}$ is consistent with the uniform distribution with $\chi^2/\rm ndf = 12.8/10$, corresponding to the spin-0 hypothesis for the $\KpKm$ system in the $\jpsi \KpKm$ final state.

\subsection{Search for the  $\Bdb \to \jpsi \phi$ decay }
The branching fraction of $\Bdb \to \jpsi \phi$ is expected to be significantly suppressed, as the  decay process $\Bdb \to \jpsi \phi$ involves hadronic final state interactions at leading order.   Here we search for the process by adding the $\phi$ resonance to the default Dalitz model. A Breit-Wigner function is used to model the $\phi$ lineshape with mass $1019.455\pm0.020$~\mev and width $4.26\pm0.04$~\mev~\cite{PDG}. The mass resolution is $\approx0.7$ MeV at the $\phi$ mass peak, which is added to the fit model by increasing the Breit-Wigner width of the $\phi$ to 4.59 MeV. We do not find any evidence for the $\phi$ resonance. The best fit value for the $\phi$ fraction, constrained to be non-negative,  is 0\%. The corresponding upper limit  at 90\% CL is determined by generating 2000 experiments from the results of the fit with the $\phi$ resonance, where the correlations of the fitted parameters are also taken into account.  The 90\% CL upper limit on the $\phi$ fraction, defined as the fraction value that exceeds the results observed in  90\% of the experiments, is 3.3\%.  
The branching fraction upper limit is then the product of the fit fraction upper limit  and the total branching fraction for  $\Bdb\to\jpsi\KpKm$. 


\section{Branching fractions}
Branching fractions are measured using the $\Bm\to\jpsi\Km$ decay mode as normalization. This decay mode, in addition to having a well-measured branching fraction, has the advantage of having two muons in the final state and being collected through  the same triggers as the \Bdb decays. The branching fractions are calculated using
\begin{equation}
\mathcal{B}(\Bdb \to \jpsi \KpKm)=\frac{N_{\Bdb}/\epsilon_{\Bdb}}{N_{\Bm}/\epsilon_{\Bm}}\times \mathcal{B}(\Bm \to \jpsi \Km),\label{eq:br}
\end{equation}
where  $N$ represents the observed yield of the decay of interest and $\epsilon$ corresponds to  the overall efficiency. We form an average of $\mathcal{B}(\Bm \to \jpsi \Km)= (10.18\pm 0.42) \times 10^{-4}$ using the  Belle~~\cite{Abe:2002rc} and BaBar~\cite{Aubert:2004rz} measurements, corrected to take into account different rates of $\Bp\Bm$ and $\Bd\Bdb$ pair production from $\Upsilon(4S)$ using $\frac{\Gamma(\Bp\Bm)}{\Gamma(\Bd\Bdb)}=1.055\pm0.025$~\cite{PDG}. 	

The detection efficiency is obtained from simulations and is  a product of the geometrical acceptance of the detector, the combined reconstruction and selection efficiency and the trigger efficiency. Since the efficiency to detect the $\jpsi \KpKm$ final state is not uniform across the Dalitz plane, the efficiency is averaged according to the default Dalitz model. Small corrections are applied to account for differences between the simulation and the data.
To ensure that the $p$ and \pt distributions of the generated $B$ meson are correct we weight the simulated samples to match the distributions of the corresponding data.  Since the normalization channel has a different number of charged tracks than the signal channel,  we weight the simulated samples with the tracking efficiency ratio by comparing the data and simulations in the track's $p$ and $\pt$ bins. Finally, we weight the simulations according to the kaon identification efficiency.
The average of the weights is assigned as a correction factor. Multiplying the detection efficiencies and correction factors gives the overall efficiencies  $(0.820\pm 0.012)\%$ and $(2.782\pm0.047)\%$ for $\Bdb \to \jpsi \KpKm$ and $\Bm\to\jpsi\Km$, respectively.

The resulting branching fraction is
\begin{equation*}
\mathcal{B}(\Bdb \to \jpsi \Kp\Km) = (2.53\pm 0.31 \pm 0.19)\times 10^{-6},
\end{equation*}
where the first uncertainty is statistical and the second is systematic. The systematic uncertainties are discussed  in Section~\ref{sec:sys}. This branching fraction has not been measured previously. 

The product branching fraction of the $a_0(980)$ resonance mode is measured for the first time, yielding 
\begin{equation*}
\mathcal{B}(\Bdb \to \jpsi a_0(980),~a_0(980) \to \KpKm)=(4.70\pm 3.31\pm0.72)\times10^{-7},
\end{equation*}
calculated by multiplying the corresponding fit fraction from the default model and the total branching fraction of the $\Bdb\to\jpsi\KpKm$ decay. The difference between the default and alternate model is assigned as a systematic uncertainty. The $a_0(980)$ resonance has a statistical significance of 3.9 standard deviations, showing evidence of the existence of $\Bdb\to\jpsi a_0(980)$ with $a_0(980)\to \KpKm$. Since the significance is less than five standard deviations, we also quote an upper limit on the branching fraction,
\begin{equation*}
\mathcal{B}(\Bdb \to \jpsi a_0(980),~a_0(980) \to \KpKm) <9.0\times 10^{-7}
\end{equation*}
at 90\%  CL.  The limit is calculated assuming a Gaussian distribution as the central value plus 1.28 times  the statistical and systematic uncertainties added in quadrature.


The upper limit of $\mathcal{B}(\Bdb \to \jpsi \phi)$ is determined to be
\begin{equation*}
\mathcal{B}(\Bdb \to \jpsi \phi) < 1.9 \times 10^{-7}
\end{equation*}
at 90\% CL, where the branching fraction $\mathcal{B}(\phi\to\KpKm)=(48.9\pm0.5)\%$ is used and the systematic uncertainties on the branching fraction of  $\Bdb\to\jpsi\KpKm$ are included. The limit  improves upon the previous best  limit of $<9.4\times 10^{-7}$ at 90\% CL, given by the \belle collaboration~\cite{Liu:2008bta}. 
According to a theoretical calculation based on $\omega-\phi$ mixing (see Appendix~\ref{sec:mixing}) the branching fraction of $\Bdb\to \jpsi\phi$ is expected to be $(1.0\pm0.3)\times10^{-7}$, which is consistent with our limit.
\section{Systematic uncertainties}
\label{sec:sys}
The systematic uncertainties on the branching fractions are estimated from the  contributions listed in Table~\ref{tab:sys_br}. Since the branching fractions are measured with respect to the $\Bm\to \jpsi\Km$ mode, which has a different number of charged tracks than the decays of interest, a 1\% systematic uncertainty is assigned due to differences in the tracking performance between data and simulation. A 2\% uncertainty is assigned  for  the decay in flight, large multiple scatterings and hadronic interactions of the additional kaon. 

\begin{table}[h!t!p!]
\centering
\caption{\small Relative systematic uncertainties on branching fractions~(\%).}
\vspace{0.2cm}
\begin{tabular}{lcc}
\hline
Source of uncertainty & $\jpsi \Kp\Km$ & $\jpsi a_0(980)$  \\
\hline  
Tracking efficiency & 1.0&1.0 \\
Material and physical effects &2.0 & 2.0\\
PID efficiency & 1.0 &1.0 \\
$\Bdb$ $p$ and $\pt$ distributions & 0.5 &0.5\\
$\Bm$ $p$ and $\pt$ distributions & 0.5 &0.5\\
Simulation sample size & 0.6&0.6\\
Background modeling& 5.7 & 5.7\\
$\mathcal{B}(\Bm\to \jpsi\Km)$&4.1&4.1\\
Alternate model&-&13.4\\
\hline
Total &7.5&15.4\\
\hline
       
\end{tabular}
\label{tab:sys_br}
\end{table}

Small uncertainties are introduced if the simulation does not have the correct $\B$ meson kinematic distributions. The measurement is relatively insensitive  to any of these differences in the $\B$ meson $p$ and $\pt$ distributions since we are measuring the relative rates. By varying the $p$ and $\pt$ distributions we see  a maximum difference of 0.5\%. There is a 1\% systematic uncertainty assigned for the relative particle identification efficiencies. 
We find a 5.7\% difference in the \Bdb signal yield when the shape of the combinatorial background is changed from a linear to a parabolic function. In addition, the difference of the $a_0(980)$ fraction between the default and alternate fit models is assigned as a systematic uncertainty for the $\mathcal{B}(\Bdb\to \jpsi a_0(980),~a_0(980)\to\KpKm)$ upper limit.
The total systematic uncertainty is obtained by adding each source of systematic uncertainty in quadrature as they are assumed to be uncorrelated.

\section{Conclusions}
We report the first observation  of the $\Bdb \to \jpsi \KpKm$ decay. The branching fraction is determined to be
\begin{equation*}
\mathcal{B}(\Bdb \to \jpsi \Kp\Km) = (2.53\pm 0.31 \pm 0.19)\times 10^{-6},
\end{equation*}
where the first uncertainty is statistical and the second is systematic.  The resonant structure of the decay is studied using a modified Dalitz plot analysis where we  include the helicity angle of the $\jpsi$. The decay is dominated by an S-wave in the $\KpKm$ system. The product branching fraction of the $a_0(980)$ resonance mode is measured to be
\begin{equation*}
\mathcal{B}(\Bdb \to \jpsi a_0(980),~a_0(980) \to \KpKm)=(4.70\pm 3.31\pm0.72)\times10^{-7},
\end{equation*}
which corresponds to a 90\% CL  upper limit of $\mathcal{B}(\Bdb \to \jpsi a_0(980),~a_0(980) \to \KpKm)<9.0\times10^{-7}$.
We also set an upper limit of  $\mathcal{B}(\Bdb \to \jpsi \phi) < 1.9 \times 10^{-7}$ at the 90\% CL. This result represents an improvement of about a factor of five with respect to the previous best measurement~\cite{Liu:2008bta}.

\section*{Acknowledgements}
 
\noindent We express our gratitude to our colleagues in the CERN
accelerator departments for the excellent performance of the LHC. We
thank the technical and administrative staff at the LHCb
institutes. We acknowledge support from CERN and from the national
agencies: CAPES, CNPq, FAPERJ and FINEP (Brazil); NSFC (China);
CNRS/IN2P3 and Region Auvergne (France); BMBF, DFG, HGF and MPG
(Germany); SFI (Ireland); INFN (Italy); FOM and NWO (The Netherlands);
SCSR (Poland); MEN/IFA (Romania); MinES, Rosatom, RFBR and NRC
``Kurchatov Institute'' (Russia); MinECo, XuntaGal and GENCAT (Spain);
SNSF and SER (Switzerland); NAS Ukraine (Ukraine); STFC (United
Kingdom); NSF (USA). We also acknowledge the support received from the
ERC under FP7. The Tier1 computing centres are supported by IN2P3
(France), KIT and BMBF (Germany), INFN (Italy), NWO and SURF (The
Netherlands), PIC (Spain), GridPP (United Kingdom). We are thankful
for the computing resources put at our disposal by Yandex LLC
(Russia), as well as to the communities behind the multiple open
source software packages that we depend on.
  


\section*{Appendix}

\appendix

\section{\mbox{\boldmath$\omega-\phi$} mixing}
\label{sec:mixing}
In Ref.~\cite{Gronau:2008kk}, Gronau and Rosner pointed out that the decay $\Bdb\to\jpsi\phi$ can proceed via $\omega-\phi$ mixing and predicted ${\cal{B}}(\Bdb\to\jpsi\phi)= (1.8\pm 0.3)\times10^{-7}$, using ${\cal{B}}(\Bdb\to\jpsi\rho)= (2.7\pm0.4)\times10^{-5}$~\cite{PDG} as there was no measurement of ${\cal{B}}(\Bdb\to\jpsi\omega)$ available. Recently  \lhcb  has measured ${\cal{B}}(\Bdb\to\jpsi\omega)=(2.41\pm0.52\,^{+0.41}_{-0.50})\times10^{-5}$~\cite{LHCb:2012cw}, which can be used to update the prediction.

The mixing $\omega-\phi$ is parametrized by a $2\times 2$ rotation matrix characterized by the angle $\delta_m$ such that the physical $\omega$ and $\phi$ are related to the ideally mixed states $\omega^I\equiv\frac{1}{\sqrt{2}}(\uubar+\ddbar)$ and $\phi^I\equiv \ssbar$, giving
\begin{eqnarray}
  \label{eq:mix}
 \omega&=&\;\;\;\cos\delta_m\omega^I+\sin\delta_m\phi^I\nonumber\\
  \phi&=&-\sin\delta_m\omega^I+\cos\delta_m\phi^I.
\end{eqnarray}
This implies 
\begin{equation}
{\cal{B}}(\Bdb\to\jpsi\phi)=\tan^2\delta_m{\cal{B}}(\Bdb\to\jpsi\omega) \Phi,
\end{equation}
where $\Phi$ represents the ratio of phase spaces  between the processes $\Bdb\to\jpsi\phi$ and $\Bdb\to \jpsi \omega$.

A simplified analysis~\cite{Benayoun:1999fv} implies a mixing angle of $\delta_m=(3.34\pm0.17)^{\circ}$, while allowing an energy dependent $\delta_m$ gives  values of  $2.75^{\circ}$ at the $\omega$ mass and $3.84^{\circ}$ at the $\phi$ mass~\cite{Benayoun:2009im}. Using the recent \lhcb value of 
${\cal{B}}(\Bdb\to\jpsi\omega)$
and $3.84^{\circ}$ for $\delta_m$, we estimate  the following value 
\begin{equation}
{\cal{B}}(\Bdb\to\jpsi\phi)= (1.0\pm0.3)\times10^{-7}.
\end{equation}

\input{B02JpsiKK.bbl}

\end{document}

%% file: lhcb.tex
\def\lhcb {LHCb\xspace}
\def\ux85 {UX85\xspace}

\def\babar  {BaBar\xspace}
\def\belle  {Belle\xspace}

\ifthenelse{\boolean{uprightparticles}}%
{

 \def\Ppi         {\ensuremath{\uppi}\xspace}

 \def\Ppsi        {\ensuremath{\uppsi}\xspace}

 \def\PDelta      {\ensuremath{\Delta}\xspace}                 
 \def\PXi      {\ensuremath{\Xi}\xspace}                 
 \def\PLambda      {\ensuremath{\Lambda}\xspace}                 
 \def\PSigma      {\ensuremath{\Sigma}\xspace}                 
 \def\POmega      {\ensuremath{\Omega}\xspace}                 
 \def\PUpsilon      {\ensuremath{\Upsilon}\xspace}                 
 
 \def\PB      {\ensuremath{\mathrm{B}}\xspace}                 
                  
 \def\PD      {\ensuremath{\mathrm{D}}\xspace}

 \def\PJ      {\ensuremath{\mathrm{J}}\xspace}                 
 \def\PK      {\ensuremath{\mathrm{K}}\xspace}

 \def\Pb      {\ensuremath{\mathrm{b}}\xspace}                 
 \def\Pc      {\ensuremath{\mathrm{c}}\xspace}                 
 \def\Pd      {\ensuremath{\mathrm{d}}\xspace}

 \def\Pi      {\ensuremath{\mathrm{i}}\xspace}

 \def\Pp      {\ensuremath{\mathrm{p}}\xspace}

 \def\Ps      {\ensuremath{\mathrm{s}}\xspace}                 
                  
 \def\Pu      {\ensuremath{\mathrm{u}}\xspace}

}
{

 \def\Ppi         {\ensuremath{\pi}\xspace}

 \def\Ppsi        {\ensuremath{\psi}\xspace}                 
                  
 \mathchardef\PDelta="7101
 \mathchardef\PXi="7104
 \mathchardef\PLambda="7103
 \mathchardef\PSigma="7106
 \mathchardef\POmega="710A
 \mathchardef\PUpsilon="7107
                  
 \def\PB      {\ensuremath{B}\xspace}                 
                  
 \def\PD      {\ensuremath{D}\xspace}

 \def\PJ      {\ensuremath{J}\xspace}                 
 \def\PK      {\ensuremath{K}\xspace}

 \def\Pb      {\ensuremath{b}\xspace}                 
 \def\Pc      {\ensuremath{c}\xspace}                 
 \def\Pd      {\ensuremath{d}\xspace}

 \def\Pi      {\ensuremath{i}\xspace}

 \def\Pp      {\ensuremath{p}\xspace}

 \def\Ps      {\ensuremath{s}\xspace}                 
                  
 \def\Pu      {\ensuremath{u}\xspace}

}

\def\uquark    {\ensuremath{\Pu}\xspace}
\def\uquarkbar {\ensuremath{\overline \uquark}\xspace}
\def\uubar     {\ensuremath{\uquark\uquarkbar}\xspace}
\def\dquark    {\ensuremath{\Pd}\xspace}
\def\dquarkbar {\ensuremath{\overline \dquark}\xspace}
\def\ddbar     {\ensuremath{\dquark\dquarkbar}\xspace}
\def\squark    {\ensuremath{\Ps}\xspace}
\def\squarkbar {\ensuremath{\overline \squark}\xspace}
\def\ssbar     {\ensuremath{\squark\squarkbar}\xspace}
\def\cquark    {\ensuremath{\Pc}\xspace}
\def\cquarkbar {\ensuremath{\overline \cquark}\xspace}
\def\ccbar     {\ensuremath{\cquark\cquarkbar}\xspace}
\def\bquark    {\ensuremath{\Pb}\xspace}

\def\pion  {\ensuremath{\Ppi}\xspace}

\def\pip   {\ensuremath{\pion^+}\xspace}
\def\pim   {\ensuremath{\pion^-}\xspace}

\def\kaon  {\ensuremath{\PK}\xspace}
  \def\Kbar  {\kern 0.2em\overline{\kern -0.2em \PK}{}\xspace}
\def\Kb    {\ensuremath{\Kbar}\xspace}
\def\Kz    {\ensuremath{\kaon^0}\xspace}
\def\Kzb   {\ensuremath{\Kbar^0}\xspace}
\def\KzKzb {\ensuremath{\Kz \kern -0.16em \Kzb}\xspace}
\def\Kp    {\ensuremath{\kaon^+}\xspace}
\def\Km    {\ensuremath{\kaon^-}\xspace}

\def\KpKm  {\ensuremath{\Kp \kern -0.16em \Km}\xspace}

\def\Kstarzb {\ensuremath{\Kbar^{*0}}\xspace}

\def\Kstarb  {\ensuremath{\Kbar^*}\xspace}

  \def\Dbar    {\kern 0.2em\overline{\kern -0.2em \PD}{}\xspace}
\def\D       {\ensuremath{\PD}\xspace}

\def\Dz      {\ensuremath{\D^0}\xspace}
\def\Dzb     {\ensuremath{\Dbar^0}\xspace}
\def\DzDzb   {\ensuremath{\Dz {\kern -0.16em \Dzb}}\xspace}
\def\Dp      {\ensuremath{\D^+}\xspace}
\def\Dm      {\ensuremath{\D^-}\xspace}

\def\DpDm    {\ensuremath{\Dp {\kern -0.16em \Dm}}\xspace}

\def\B       {\ensuremath{\PB}\xspace}
  \def\Bbar    {\kern 0.18em\overline{\kern -0.18em \PB}{}\xspace}

\def\Bu      {\ensuremath{\B^+}\xspace}
\def\Bub     {\ensuremath{\B^-}\xspace}
\def\Bp      {\ensuremath{\Bu}\xspace}
\def\Bm      {\ensuremath{\Bub}\xspace}

\def\Bd      {\ensuremath{\B^0}\xspace}

\def\Bsb     {\ensuremath{\Bbar^0_\squark}\xspace}
\def\Bdb     {\ensuremath{\Bbar^0}\xspace}

\def\jpsi     {\ensuremath{{\PJ\mskip -3mu/\mskip -2mu\Ppsi\mskip 2mu}}\xspace}

  \def\Y#1S{\ensuremath{\PUpsilon{(#1S)}}\xspace}

\def\proton      {\ensuremath{\Pp}\xspace}
\def\antiproton  {\ensuremath{\overline \proton}\xspace}

\def\L {\ensuremath{\PLambda}\xspace}
\def\Lbar {\ensuremath{\kern 0.1em\overline{\kern -0.1em\Lambda\kern -0.05em}\kern 0.05em{}}\xspace}

\def\Lb      {\ensuremath{\L^0_\bquark}\xspace}

\def\to                 {\ensuremath{\rightarrow}\xspace}

\def\CP                {\ensuremath{C\!P}\xspace}

\def\AT#1     {\ensuremath{A_{\mathrm{T}}^{#1}}\xspace}           

\def\C#1      {\ensuremath{\mathcal{C}_{#1}}\xspace}                       
\def\Cp#1     {\ensuremath{\mathcal{C}_{#1}^{'}}\xspace}                    
\def\Ceff#1   {\ensuremath{\mathcal{C}_{#1}^{\mathrm{(eff)}}}\xspace}        
\def\Cpeff#1  {\ensuremath{\mathcal{C}_{#1}^{'\mathrm{(eff)}}}\xspace}       
\def\Ope#1    {\ensuremath{\mathcal{O}_{#1}}\xspace}                       
\def\Opep#1   {\ensuremath{\mathcal{O}_{#1}^{'}}\xspace}                    

\newcommand{\tev}{\ensuremath{\mathrm{\,Te\kern -0.1em V}}\xspace}
\newcommand{\gev}{\ensuremath{\mathrm{\,Ge\kern -0.1em V}}\xspace}
\newcommand{\mev}{\ensuremath{\mathrm{\,Me\kern -0.1em V}}\xspace}
\newcommand{\kev}{\ensuremath{\mathrm{\,ke\kern -0.1em V}}\xspace}
\newcommand{\ev}{\ensuremath{\mathrm{\,e\kern -0.1em V}}\xspace}
\newcommand{\gevc}{\ensuremath{{\mathrm{\,Ge\kern -0.1em V\!/}c}}\xspace}
\newcommand{\mevc}{\ensuremath{{\mathrm{\,Me\kern -0.1em V\!/}c}}\xspace}
\newcommand{\gevcc}{\ensuremath{{\mathrm{\,Ge\kern -0.1em V\!/}c^2}}\xspace}
\newcommand{\gevgevcccc}{\ensuremath{{\mathrm{\,Ge\kern -0.1em V^2\!/}c^4}}\xspace}
\newcommand{\mevcc}{\ensuremath{{\mathrm{\,Me\kern -0.1em V\!/}c^2}}\xspace}

\def\mum  {\ensuremath{\,\upmu\rm m}\xspace}

\def\gsim{{~\raise.15em\hbox{$>$}\kern-.85em
          \lower.35em\hbox{$\sim$}~}\xspace}
\def\lsim{{~\raise.15em\hbox{$<$}\kern-.85em
          \lower.35em\hbox{$\sim$}~}\xspace}

\def\pt         {\mbox{$p_{\rm T}$}\xspace}

\def\evtgen     {\mbox{\textsc{EvtGen}}\xspace}
\def\pythia     {\mbox{\textsc{Pythia}}\xspace}

\def\geant      {\mbox{\textsc{Geant4}}\xspace}

\def\gauss      {\mbox{\textsc{Gauss}}\xspace}

\def\photos     {\mbox{\textsc{Photos}}\xspace}

\def\tell1  {TELL1\xspace}
\def\ukl1   {UKL1\xspace}



%% file: lhcb_author_list.tex
\centerline{\large\bf LHCb collaboration}
\begin{flushleft}
\small
R.~Aaij$^{40}$, 
B.~Adeva$^{36}$, 
M.~Adinolfi$^{45}$, 
C.~Adrover$^{6}$, 
A.~Affolder$^{51}$, 
Z.~Ajaltouni$^{5}$, 
J.~Albrecht$^{9}$, 
F.~Alessio$^{37}$, 
M.~Alexander$^{50}$, 
S.~Ali$^{40}$
% ...

{\footnotesize \it
% ...
$ ^{5}$Clermont Universit\'{e}, Universit\'{e} Blaise Pascal, CNRS/IN2P3, LPC, Clermont-Ferrand, France\\
$ ^{6}$CPPM, Aix-Marseille Universit\'{e}, CNRS/IN2P3, Marseille, France\\
% ...
$ ^{9}$Fakult\"{a}t Physik, Technische Universit\"{a}t Dortmund, Dortmund, Germany\\
% ...
$ ^{36}$Universidad de Santiago de Compostela, Santiago de Compostela, Spain\\
$ ^{37}$European Organization for Nuclear Research (CERN), Geneva, Switzerland\\
% ...
$ ^{40}$Nikhef National Institute for Subatomic Physics, Amsterdam, The Netherlands\\
% ...
$ ^{45}$H.H. Wills Physics Laboratory, University of Bristol, Bristol, United Kingdom\\
% ...
$ ^{50}$School of Physics and Astronomy, University of Glasgow, Glasgow, United Kingdom\\
$ ^{51}$Oliver Lodge Laboratory, University of Liverpool, Liverpool, United Kingdom\\
% ...
}
\end{flushleft}

%% file: B02JpsiKK.bbl
\ifx\mcitethebibliography\mciteundefinedmacro
\PackageError{LHCb.bst}{mciteplus.sty has not been loaded}
{This bibstyle requires the use of the mciteplus package.}\fi
\providecommand{\href}[2]{#2}